# Excitation and reception of magnetostatic surface spin waves in thin conducting ferromagnetic films by coplanar microwave antennas. Part I: Theory

Charles Weiss[1], Matthieu Bailleul[2] and Mikhail Kostylev[1*]

1. Department of Physics and Astrophysics, The University of Western Australia, Crawley, WA 6009, Australia
2. Institut de Physique et Chimie des Matériaux de Strasbourg, 23 rue du Loess, BP 43, 67034 Strasbourg Cedex 2, France

## Abstract

A fully self-consistent model for the excitation and reception of magnetostatic surface waves in thin ferromagnetic films by a set of coplanar antennas was developed and implemented numerically. The model assumes that the ferromagnetic film is highly conducting and is interfaced with non-magnetic metallic films, but is also suitable for modeling magneto-insulating films. Perpendicular magnetic anisotropy and Dzyaloshinskii-Moriya interaction can be included at both interfaces of the ferromagnetic layer. The model calculates the coupling impedances between the different strips constituting the coplanar antennas. In some situations, this leads to a frequency non-reciprocity between counter-propagating waves even in the case of no asymmetry in the spin-wave dispersion relation. Several intermediate results of the model were checked numerically and the final output of the model, given as the scattering parameters, $S_{11}$, $S_{12}$, and $S_{21}$ of the antenna system, were in good agreement with previous experimental studies.

## 1. Introduction

Magnonics is a field of research and technology that studies and exploits microwave magnetization dynamics in the form of traveling and standing spin waves, or magnons in the quantum language. Multiple microwave signal processing devices based on magnons have been suggested [1], as well as applications for sensing fields [2,3], substances [4,5], and nano-objects [6]. Furthermore, traveling-spin-wave-based logic gates have been implemented, potential applications of traveling spin waves in Quantum Computing and Information [7–9] have been discussed, and neuromorphic reservoir computing using traveling spin waves has been demonstrated [10].

Traveling spin waves can also be useful for characterizing magnetic materials. For instance, recently, it has been suggested that measuring spin-wave dispersion in ferromagnetic films with the interface Dzyaloshinskii-Moriya interaction (iDMI) is the most direct way of extracting the value of the Dzyaloshinskii constant for these materials [11–17]. The films with iDMI are prospective candidates for applications in new magnetic memory and logic based on skyrmions [18]. It is very important to be able to measure the value of the Dzyaloshinskii constant for the materials that can support the formation of skyrmions, and measurements of spin-wave dispersion provide this opportunity.

---

[*] Corresponding author, mikhail.kostylev@uwa.edu.au

The spin-wave delay line (SWDL) [19,20] is the basic traveling magnonic device. It consists of a ferromagnetic film and two parallel stripline spin-wave transducers (antennas) on its top. One antenna is used as the input transducer that converts an input microwave signal into a signal carried by a spin wave in the ferromagnetic film. The second (output) antenna converts the spin-wave signal back into a microwave signal in the output transmission line.

Most of the applications of the traveling spin waves listed above employ SWDL as the main building block. Historically, the first spin-wave devices employing ferromagnetic films were of macroscopic sizes – the propagation path of spin waves between the two antennas was several millimeters long [20]. Those devices were based on low-magnetic-moment yttrium-iron garnet (YIG) films with thicknesses in the micrometer range ("thick YIG films"). Later on, microscopic spin-wave devices employing films with high magnetic moments made of ferromagnetic metals were proposed [19]. These films are much thinner – their typical thickness is 30 nm. In particular, the microscopic SWDL structures were used to observe the Doppler shift of spin waves [21], and to measure the spin-wave dispersion in ferromagnetic films with iDMI [14,17,22–24].

An important peculiarity of spin waves in magnetic films with thicknesses in the nanometer range is that only the magnetostatic surface spin wave (MSSW), which propagates in an in-plane magnetized film and perpendicular to the field direction, is useful for applications, as explained in Section 1 of the online supplementary materials [25].

One more peculiarity of the microscopic delay lines is that a specific type of microwave transducer has to be used to ensure reasonable efficiency of excitation of spin waves by localized microwave currents in the transducers. Most macroscopic SWDL devices employ microstrip transducers. The microstrip transducer represents a section of a microstrip line shorted at its far end. A YIG film, epitaxially grown on a separate (gadolinium gallium garnet) substrate, is simply placed on top of the microstrip antenna structure, thus creating a hybrid chip. The microstrip transducers are used because they ensure the efficient excitation of spin waves in a broad range of frequencies. The maximum spin-wave excitation efficiency for the microstrip antennas corresponds to zero wavenumber $k_x=0$, where the spin-wave group velocity is maximized. (More precisely, the maximum is located close to $k_x=0$, but not exactly at $k_x=0$, because of the effect of the ground plane of the microstrip line [7]. Importantly, relatively wide microstrip transducers, typically having a width of 50 microns [9], are employed in association with the micron-thick YIG films. Transducers with such significant widths can be used because the length of the free propagation path of SWs in those films amounts to several millimeters. Microstrips of this width are mechanically strong enough to withstand the pressure of placing YIG films on top of them. Furthermore, the efficiency of coupling of the magnetization precession to the driving Oersted field of the microwave current in the microstrip transducer is large. Therefore, the characteristic impedance of the antenna off the spin-wave band does not need to be impedance-matched to the characteristic impedance of the feeding line. The effect of excitation of spin waves strongly modifies the input impedance of the antenna (within the spin-wave frequency band) [26–39], and, if the radiation impedance inserted by excitation of spin waves by the antenna is properly accounted for, almost complete conversion of the input microwave power into the power carried by spin waves in a YIG film is possible [40].

Conversely, the SW free propagation path in ferromagnetic metallic films with thicknesses in the nanometer range is just several microns [21] or several tens of microns [41]. If the antenna is much wider than the free propagation path, instead of driving the traveling-spin-wave dynamics, the forced magnetization dynamics of the near field of the antenna [41] is excited that can be treated as the excitation of the ferromagnetic resonance (FMR) in the film [42]. This effect is largely used for characterizing ferromagnetic films and is called "Broadband FMR". First when the antenna width becomes comparable to the free propagation path, does the traveling-spin-wave contribution to the precessional motion of magnetization appear [43,44].

Therefore, to excite a traveling wave that can leave the area of the excitation transducer, the transducer width should be much smaller than the free propagation path – that is from hundreds of nanometers to a couple of microns. This excludes the hybrid-chip approach for microscopic devices. The method that was found to work is to integrate the antennas on the surface of the film with the help of lithography techniques [19].

The antenna geometry that proved to be able to excite spin waves efficiently in this situation is a modification of a coplanar line (CPL) [14,17,19,21–23,41]. The standard CPL is the so-called coplanar waveguide, characterized by ground lines of infinite width [45]. The CPL used as the microscopic antenna of spin waves has ground lines of finite width. The width is close to the width of the central lead of the line (called the signal line). One important advantage of this geometry is that it is perfectly two-dimensional – the ground and the signal lines lie in the same plane, which simplifies significantly the fabrication.

One more peculiarity of spin-wave excitation by the CPL is that the maximum excitation efficiency corresponds to a much larger wavenumber than in the case of microstrip antennas. This allows creating narrow-band band-pass filters based on SWDL lines employing the CPL transducers [21].

The comprehensive self-consistent theory of excitation of spin waves in thick YIG films by the standard CPL antennas having ground lines of infinite width was developed by Vugalter et al. [30,32,36]. Contrary to the very first theories [27–29], the self-consistent method does not use any assumption for the form of the distribution of the microwave current density across the strips of a stripline antenna. Instead, the distribution is obtained by solving an integral equation for the current density. Then, it is used to calculate the microwave characteristics of the antenna [46]. Numerical calculations with this theory could be carried out only for a couple of special cases, for instance, for negligibly small wavenumbers. This was because of the complexity of the derived equations. Furthermore, the magnetic film was considered perfectly insulating. This is suitable for YIG films, but is not acceptable for the ferromagnetic metallic films because of their very large conductivities.

The first goal of the present work is to develop a comprehensive self-consistent theory of operation of a microscopic spin-wave device that represents a ferromagnetic film with two coplanar antennas on its top. We will treat separately the processes of excitation of a spin wave by the input (excitation) antenna and the reception by the output (receiving) antenna of a signal carried by the spin wave. We will do this by developing separate self-consistent theories of excitation and reflection. In addition, our theories will:
- account for the high conductivity of the ferromagnetic film, which is important for ferromagnetic metals, such as iron, cobalt, nickel, and their alloys;
- consider a ferromagnetic layer as being a part of a layer stack consisting of metallic non-magnetic layers with the ferromagnetic layer sandwiched in-between;
- include the contribution of the inhomogeneous exchange interaction to the SW energy;
- make use of exchange boundary conditions that include perpendicular magnetic anisotropy (PMA) and Dzyaloshinskii-Moriya interaction (iDMI) at the interfaces of the non-magnetic layers with the ferromagnetic one;
- include the intrinsic Ohmic resistance of the antenna strips directly into the integral equations of a self-consistent theory (earlier theories did it first at the Telegrapher Equations stage which is not a rigorous approach);
- allow the ground lines of the coplanar antennas to have arbitrary finite widths such that off the spin-wave band our solution will reduce to the characteristic impedance and propagation constant for the general case of a backed coplanar line with an arbitrary width of ground lines and not loaded by a magnetic layer (the multilayer film of finite thickness will then act as the backing plate);
- cast the solution of the integral equation that we will obtain into an inductive-impedance tensor for the coplanar antennas;
- include the effect of non-reciprocity of spin-wave excitation and reception by the coplanar antennas into the Telegrapher Equations that we will derive. This will be enabled by the introduction of the inductive impedance tensor. As a result, the system of Telegrapher Equations will consist of five equations, contrary to earlier works that operated with the standard system of two equations (see e.g. a review article [36]).

This extensive set of features has never been combined in any earlier theory. Furthermore, earlier works, e.g. [36], focused on obtaining approximate analytical solutions for integral equations. Conversely, we aim to cast the integral equations in a form that will allow one to easily solve the equations numerically, to be able to calculate the *S*-parameters of the spin-wave device with high precision. Solutions of the Telegrapher Equations that we will obtain will allow us to easily convert the numerical solutions of the integral equations into values of *S*-parameters. Because the solutions of the Telegrapher Equations are quasi-analytical, this step will result in a fast numerical computation of those parameters. At the end of the day, this combination of features will allow the conversion of the theory into a program code that will be easy to use.

Note that despite the inclusion of the conductivity of both magnetic and non-magnetic layers, our model will also be suitable for insulating magnetic films of any thickness, including recently developed nanometer-range thick YIG films [47]. To this end, one will need to set the film conductivity to a negligibly small value in the final equations of the theory.

The second goal of this work is to use the constructed theory to investigate the details of the operation of the spin-wave device. We will focus on the effect of non-reciprocity of excitation and reception of spin waves by the coplanar antennas, look at the mechanism of formation of the transmission ($S_{21}$ and $S_{12}$) and reflection ($S_{11}$ and $S_{22}$) frequency bands of the device and explore the effects of PMA, iDMI and other magnetic parameters of the ferromagnetic layer on those characteristics. We will also discuss to which extent the approximations employed in earlier treatments can be justified for practical purposes.

Following these aims, our paper is organized as follows. Section 1 is the Introduction. Section 2 is devoted to the construction of the theory. First, we set the problem (Sub-section 2.1). The next step (Sub-section 2.2) is the derivation of the integral equations and their solution in the form of the inductive-impedance tensor. Sub-section 2.3 is devoted to the derivation of the Telegrapher Equations and their solution in the form of the input impedance of the input spin-wave antenna. Ultimately, the input impedance is converted into the $S_{11}$ parameter for the antenna. The next sub-section (2.4) deals specifically with developing a model for the reception of a spin-wave signal by the output antenna. The ultimate result of this section is an expression for $S_{21}$.

Section 3 reports on the details of a numerical model we developed based on the constructed theory and results of numerical simulations with the model. We use the model to investigate the details of the operation of the device, per the second goal of our work (see above). Some details of this investigation are placed into the online supplemental materials to this article [25].

Conclusions from our investigation are placed into Section 4, and more technical details of the theory are also placed into Ref. [25].

## 2. Theoretical model[†]

### 2.1. Problem setting

We consider a metallic ferromagnetic film of thickness $L$ that is homogeneous throughout its volume. From each side, it is interfaced with a non-magnetic (relative permittivity, $\mu=1$) metallic layer. It is assumed that perpendicular magnetic anisotropy (PMA) [48] and Dzyaloshinskii-Moriya interaction [49] can be present at the interfaces. They are described by constants of the interface PMA $K_{u1}$ and $K_{u2}$, and Dzyaloshinskii constants $D_1$ and $D_2$. Here indexes 1 and 2 correspond to the interfaces with the non-magnetic metallic layers of thickness $d_1$ and $d_2$ respectively. The film stack and thicknesses are displayed in Figure 1(a).

In the first stage of the calculation, we treat the layered film as infinitely long along both in-plane axes - $x$ and $z$. In the second stage, a finite width in the direction $z$ will be assumed. The external magnetic field $H$ is applied in the positive $z$-direction. The tri-layer metallic film sits on top of a dielectric substrate. As we treat the substrate in the magnetostatic approximation ($\varepsilon=0$) [50]], the dielectric permittivity $\varepsilon$ for the substrate does not matter. A dielectric spacer of a thickness $d_s$ separates the upper metal layer $d_2$ from a coplanar line (CPL). The dielectric constant for the spacer is $\varepsilon_s$. We will need $\varepsilon_s$ to solve a separate problem of calculation of the linear capacitance for the CPL. (The definition of the linear capacitance will be provided and its physical meaning will be explained in Section 2.3.) The constant will not be used to solve the problem of modeling the spin-wave dynamics because we will employ the magnetostatic approximation to this end [50]. More precisely, we will employ a slightly more general displacement-current-free approximation (see Eq. (3) below), because we need to include the conductivities of metallic layers in the theory. Following previous treatments [36,38], we assume that the thickness (in the direction $y$) of the line conductors is negligibly small. Although this assumption is not fully realistic, it is required to avoid a very heavy fully numerical treatment. The width of the signal line (the central strip) in the direction $x$ is $w$. The width of the ground lines (the side strips) of the CPL is $w_g$. Gaps of a width $\Delta_g$ separate the signal line from the ground lines. The area above the CPL is free space with $\varepsilon=0$ and relative magnetic permittivity $\mu=1$.

The CPL represents an input antenna of spin waves. It is fed by a microwave voltage $V=1$ Volt supplied by a feeding line with a characteristic impedance of 50 Ohms. The input port of the antenna is located at $z = l_s$. The length of the antenna in the direction $z$ is $l_s$. The antenna is shorted at its far end ($z = 0$). A second CPL with the same geometry and similarly shorted at its far end is placed at a distance $l_d$ from the input antenna. (The distance is measured between the symmetry axes of the antennas. Accordingly, the edge-to-edge separation of the antennas is $l_d-(2w_g+w+2\Delta_g)$.) A schematic of both antennas can be seen in Figure 1(b). The second CPL serves as a receiving transducer of spin waves – it converts the signal carried by a spin wave back into a microwave voltage $V_{out}$ at the output antenna port (at $z = l_s$). The port is connected to a regular transmission line with the standard characteristic impedance of 50 Ohms. The transmission line carries the converted signal to its final recipient.

---
[†] The theory from this section is by M. Kostylev

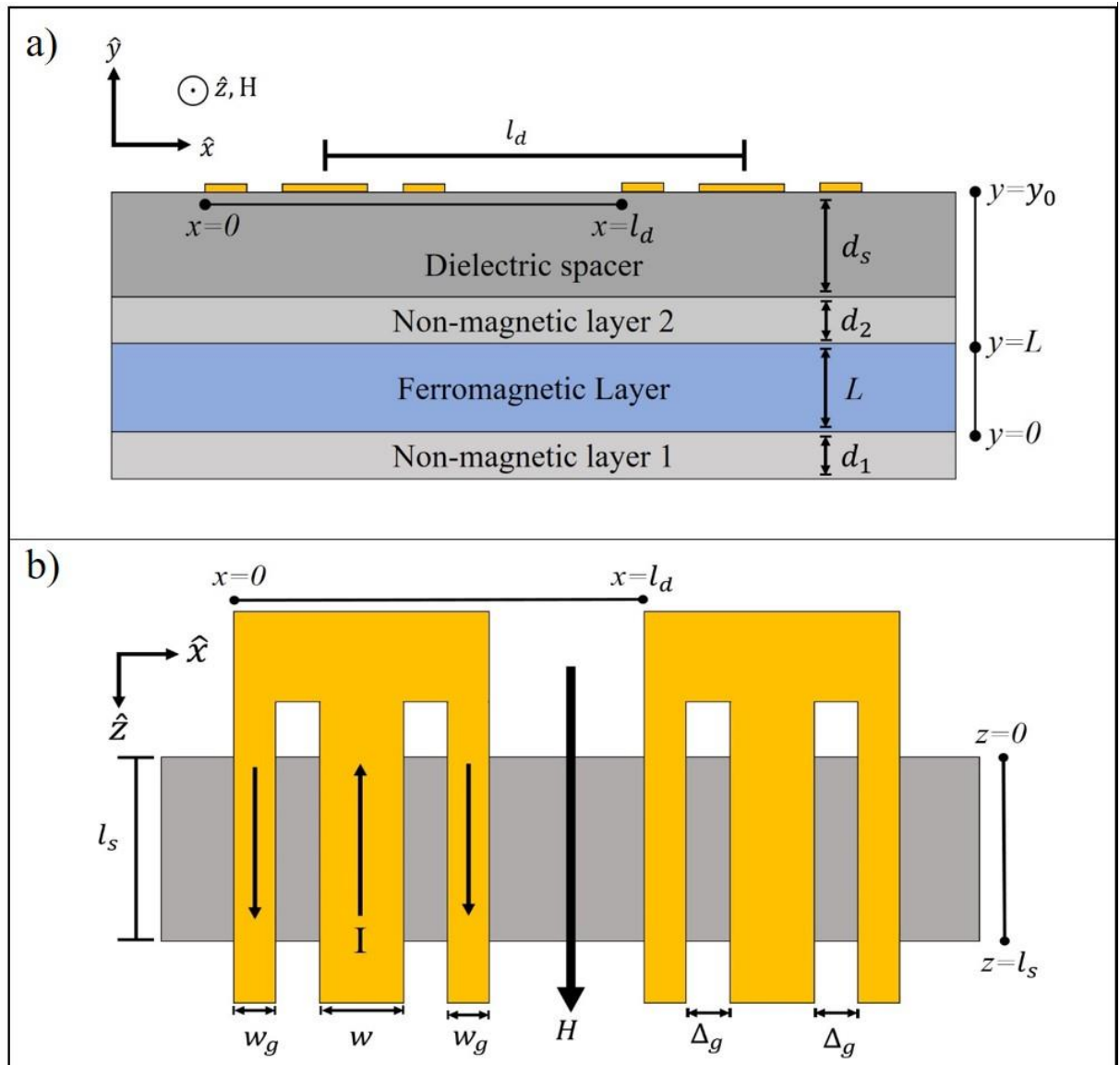

*Figure 1: a) Side view of the vertical cross-section of the multilayer with the thicknesses, $d_s$, $d_1$, $d_2$, and $L$ shown for the corresponding layers. The gold rectangles on top of the spacer correspond to the infinitely thin strips of the CPL. b) Top view of the multilayer track and CPL showing the geometry of the CPL antennas. Both antennas are identical. The x=0 point corresponds to the outer edge of the first ground line of the input antenna, with the signal and 2nd ground line placed at x>0.*

## 2.2. Antenna characteristics

In the first stage of the calculation, we deal with the electric characteristics of the antenna. The characteristics are the same for both antennas. We will cast them into the form of a matrix of linear self and mutual inductive impedances of the individual strips of a CPL. One standard approach to the calculation of the complex impedance for the striplines loaded by ferromagnetic films is by exploiting the translation invariance in the $z$-direction [34,36]. This corresponds to a quasi-static approach for the description of microwave transmission lines [51] and results in a two-dimensional problem that we will be dealing with below. The actual form of the electromagnetic field dependence on the $z$ coordinate – in the form of an electromagnetic wave traveling along the antenna - will be accounted for while employing the obtained impedance matrix for calculations of the input impedance of the input antenna, the reflection coefficient from its port ($S_{11}$) and $V_{out}$.

To describe the magnetization dynamics in the ferromagnetic layer, we employ the linearized Landau-Lifschitz equation (LLE)

$$\frac{\partial \mathbf{m}}{\partial t} = -|\gamma|\mu_0 (\mathbf{m} \times \mathbf{H} + \mathbf{M} \times \mathbf{h}_{eff}), \tag{1}$$

where $\mathrm{H} = \mathrm{H}_0 \mathrm{u}_z$, $\mathbf{M} = M_0 \mathbf{u}_z$, and $\mathbf{u}_z$ is a unit vector in the direction $z$. In Eq. (1), the dynamic magnetization $\mathbf{m}$ has only two components: $(\mathrm{m}_x, \mathrm{m}_y)$, which are perpendicular to the static magnetization M. The magnitude $M_0$ is equal to the saturation magnetization of the film $|M_0| = M_S$. The dynamic effective field $\mathrm{h}_{eff}$ consists of two parts. The effective $\mathbf{h}_{ex}$ of the inhomogeneous exchange interaction is given by [52]

$$\mathbf{h}_{ex} = \alpha \left( \frac{\partial^2}{\partial x^2} + \frac{\partial^2}{\partial y^2} \right) \mathbf{m}, \tag{2}$$

where $\alpha = 2A/(\mu_0 M_S^2)$, and $A$ is the exchange constant.

The second part is the dynamic magnetic field $\mathbf{h}$. It is the solution of the Maxwell equations in the displacement-current-free approximation:

$$\nabla \times \mathbf{h} = \sigma \mathbf{e}, \tag{3}$$

$$\nabla \cdot \mathbf{h} = -\nabla \cdot \mathbf{m}, \tag{4}$$

$$\nabla \times \mathbf{e} = -i\omega\mu_0 (\mathbf{h} + \mathbf{m}). \tag{5}$$

Here $\mathbf{e}$ is the microwave electric field, $\sigma$ is electric conductivity, $\mu_0$ is the permittivity of vacuum and $\omega$ is the frequency of the microwave driving field. Only two vector components of the dynamic magnetic field are important for the description of the linear magnetization dynamics. These are $h_x$ and $h_y$. They are perpendicular to the direction $z$ along the stripline. We denote a vector composed of these components as $\mathrm{h} = (\mathrm{h}_x, \mathrm{h}_y)$. Note that the term containing the electric permittivity is missing on the right-hand side of Eq. (3); this is consistent with the standard magnetostatic approximation for ferromagnetic materials [50]. Furthermore, for metals at microwave frequencies, the contribution of the displacement current to Eq. (3) is negligible with respect to the conductivity one [53].

We do not include the effective field of bulk anisotropy of the film in the present model. (As a bulk anisotropy we mean an anisotropy where the energy is present at any location of the film volume, contrary to PMA, where the energy is concentrated at the interfaces of the ferromagnetic layer with the non-magnetic metallic ones.) This is done for the simple reason of keeping the formalism as simple as possible. If needed, the effective anisotropy field $H_a$ can be easily introduced by employing the standard approach of the effective demagnetizing factors of anisotropy (see e.g. [54]). Note that, strictly speaking, even in the simplest case of (bulk) uniaxial anisotropy with the axis perpendicular to the film plane (which is usually referred to as "Perpendicular Magnetic Anisotropy" or "PMA") the effective demagnetizing factor approach does not reduce to the usual substitution of $M_S$ with effective saturation magnetization ($M_{eff} = M_S - H_a$) in the final formulas if inhomogeneity of magnetization dynamics in the film plane (i.e. its *x*-dependence) is included in the model, as in our case. This is seen in Figures. 3 and 4 in [54], where the effect of bulk PMA is seen not just as a mere shift of the spin-wave dispersion curves downwards, but also as a significant change in the slope of the dispersion curve (see also appendix A of [55]). Recall that surface PMA will be included in our model - as a specific boundary condition for the dynamic magnetization (see Eq. (17) below).

According to [56–58], it is convenient to approach the problem in the Fourier space:

$$\mathbf{m}, \mathbf{h}, \mathbf{e} \sim \exp(i\omega t - ikx). \tag{6}$$

Substitution of Eq. (6) into Eqs. (3)-(5) yields

$$ke_z = -\omega\mu_0(h_y + m_y), \tag{7}$$

$$\frac{\partial h_x}{\partial y} + ikh_y = -\sigma e_z, \tag{8}$$

$$-ikh_x + \frac{\partial h_y}{\partial y} = -\frac{\partial m_y}{\partial y} + ikm_x, \tag{9}$$

$$\frac{\partial e_z}{\partial y} = -i\omega\mu_0(h_x + m_x). \tag{10}$$

It is easy to verify that, similar to [57] the ansatz $h, m \sim e^{qy}$ solves the system of equations (Eqs. (1), (6)-(10)). Eqs. (7)-(10) then reduce to

$$kqh_x + iK^2 h_y - i(k^2 - K^2)m_y = 0, \tag{11}$$

$$-ikh_x + qh_y + qm_y - ikm_x = 0, \tag{12}$$

where $K^2 = k^2 + i\sigma\omega\mu_0$.

Eq. (1) then takes the form as follows:

$$h_x = \left[\frac{\omega_H}{\omega_M} + \alpha\left(k^2 - q^2\right)\right]m_x - i\frac{\omega}{\omega_M}m_y, \qquad (13)$$

$$h_y = \left[\frac{\omega_H}{\omega_M} + \alpha\left(k^2 - q^2\right)\right]m_y + i\frac{\omega}{\omega_M}m_x, \qquad (14)$$

where $\omega_H = |\gamma|\mu_0(H + i\Delta H)$, $\omega_M = \gamma M_s$, and $\Delta H = \alpha_G \omega/(|\gamma|\mu_0)$ is the magnetic loss parameter. It scales as the Gilbert magnetic damping constant $\alpha_G$. The magnetic losses have been introduced into Eqs. (13)-(14) phenomenologically, as it is known [50] that the linearized Landau-Lifshitz-Gilbert Equation reduces to the linearised Landau-Lifshitz Equation (Eq. (1)) with a complex-valued value of the applied field given by the expression above. Equations (11)-(14) form a homogeneous system of linear algebraic equations. Eliminating $h_x$ and $h_y$, the system reduces to two equations for $m_x$ and $m_y$ shown in [58].

Equating the determinant of the system of the two equations to zero yields a characteristic equation for the problem. It has the form of a bi-cubic equation (Eq. (15) in [58]). The bi-cubic equation allows an analytic solution $q_1 = \sqrt{Q_1}$, $q_2 = \sqrt{Q_2}$, $q_3 = \sqrt{Q_3}$, $q_4 = -\sqrt{Q_1}$, $q_5 = -\sqrt{Q_2}$, $q_6 = -\sqrt{Q_3}$ (see Eqs. (A3) in [58]). Note that because the coefficients of the characteristic equation are complex numbers, the roots are complex-valued.

The complete solution for the two components of **m** can be expressed in the following form

$$m_y = \sum_{i=1}^{6} M_i^y \exp(q_i y), \qquad (15)$$

$$m_x = \sum_{i=1}^{6} M_i^x \exp(q_i y). \qquad (16)$$

To determine the coefficients $M_i^x$ (or $M_i^y$), we need six boundary conditions at the film surfaces. Furthermore, to obtain a nontrivial solution, at least one of these boundary conditions should be inhomogeneous.

The first set of available boundary conditions is the exchange boundary conditions for the original Landau-Lifshitz equation. They apply to the magnetization vector [59],

$$\frac{\partial m_{x(y)}}{\partial y} \pm d_p m_{x(y)} \pm i\tilde{D}_p k m_{y(x)} = 0. \qquad (17)$$

Here $d_p$ is the interface magnetization pinning constant that originates from the interface PMA: $d_p = -K_{up}/A$ [59], $\tilde{D}_p = D_p/A$ is the contribution of the interface Dzyaloshinskii-Moriya interaction (iDMI) to the boundary condition [60], and $p=1, 2$ denotes the interface with the respective non-magnetic layer. **m** has two components - $m_x$ and $m_y$, and this exchange boundary condition applies to both interfaces of the ferromagnetic layer. This yields four exchange boundary conditions in total. Their effect on the magnetization dynamics is discussed in Section 2 of Ref. [25].

The availability of four exchange boundary conditions is still not enough to solve the problem, and we need to add two more. These extra boundary conditions are electromagnetic and originate from the solution of the boundary value problem for Eqs. (7)-(10) outside the ferromagnetic layers. Similar to [61], we may reduce the boundary value problem to boundary conditions at the interfaces of the ferromagnetic layer that include quantities for the inside of the ferromagnetic layer only. These boundary conditions read:

$$\sum_{i=1}^{6}\left[R_m C_{myi} + (S_0 + q_i R_h)C_{hxi}\right]M_i^y = 0, \qquad (18)$$

$$\sum_{i=1}^{6}\left[R_m C_{myi} + (S_1 + q_i R_h)C_{hxi}\right]M_i^y \exp(q_i L) = -S_j. \qquad (19)$$

Expressions for the quantities $R_m$, $R_h$, $C_{my}$, $C_{hx}$, $S_0$, $S_1$, and $S_j$ are given in Section 8 of Ref. [25]. The system of the six boundary conditions Eqs. (17)-(19) is inhomogeneous, as the last boundary condition has a non-vanishing right-hand side. The right-hand side originates from the Fourier transform of the microwave current density $j_{sk}$ in the input antenna. This current drives the magnetization precession **m** in the ferromagnetic layer that gives rise to the formation of traveling spin waves in the layer.

We can cast the inhomogeneous system of linear equations (Eqs. (17)-(19)) into a vector-matrix equation

$$\sum_{j=1}^{6} C_{ij} M_j^y = F_i, \qquad (20)$$

where **C** is the matrix of coefficients of the system and **F** is the vector of the right-hand side terms of the system. The formal solution of the matrix equation reads

$$\mathbf{M}^y = \mathbf{C}^{-1}\mathbf{F}, \qquad (21)$$

where $\mathbf{M}^y$ is the vector whose elements are $M_{ij}^y$.

To progress further, we now need to calculate the Fourier image of the $z$-component of the electric field $e_{zk}(y_0)$ at the antenna surface $y_0 = L + d_2 + s$. The field is induced by the excited spin wave. This expression is obtained based on the solution Eq. (21) and is shown in Section 8 of Ref. [25].

The Green's function of the electric field **e** at $y_0$ is then given by the inverse Fourier transform of the Fourier image

$$G_E(x) = \int_{-\infty}^{\infty} e_{zk}(y_0) \exp(-ikx)\,dk. \qquad (22)$$

Physically, the Green's function represents the microwave electric field of a spin wave excited by a microwave current flowing in the *z*-direction through a wire of an infinitely small *x–y* cross-section. This is seen from the fact that the Fourier transform of the current reads $j_{sk} = 1$. Accordingly, the Fourier transform of the Green's function $e_{zk}(y_0)$ has a physical meaning of the amplitude of a continuous wave of the electric field induced by magnetization dynamics, when it is driven by a harmonically-varying spatial distribution of the current density $j_{sk}$ $\exp(-ikx)$.

An electric field at a location ($x$, $y_0$) created by an arbitrary current density distribution $j(x)$ is then given by

$$E(x) = \int_w G_E(x-x')j(x')dx', \qquad (23)$$

where the integral is taken over the whole width *w* along *x* of the area where the current is flowing.

As follows from Eq. (21), the Green's function accounts for the interface PMA and DMI in the form of the appropriate exchange boundary conditions at both interfaces of the magnetic layer.

The Green's function allows one to solve the problem of excitation of spin waves by stripline antennas self-consistently (see e.g. [36]). If a current with a density $j_i(x)$ flows through the *i*-th strip of the antenna, two electric fields are created at each point of the antenna. Here strip 1 corresponds to the signal line and strips 2 and 3 correspond to the ground lines of the coplanar antenna. The first field is the electric field of the excited spin wave that is given by Eq. (23). The second field is just a linear-voltage drop due to ohmic losses in the metal of the strip

$$E_i^{Ohm}(x) = r_\square j_i(x), \qquad (24)$$

where $r_\square$ is the antenna sheet resistance measured in Ohm per square.

As shown in Section 10 of Ref. [25], for a conducting material, to which an external voltage is applied to drive the current, the sum of the two electric fields must be uniform across the widths of the CPL's strips –

$$\int_{(x_n)} G_E(x-x')j(x')dx' + E_i^{Ohm}(x) = const, \qquad (25)$$

within each individual strip, i.e. within three *x*-ranges ($x_n$), where the index $n=1,2,3$ denotes strip numbers. (Recall that we assume that the antennas are infinitely thin in the direction *y*.) *n* =1 corresponds to the signal line and *n*=2, 3 corresponds to the ground line strips at smaller, larger *x*-values, respectively (see Figure 1). This condition yields a system of three integral equations that govern the microwave current density $j_i(x)$ in the strips

$$E_i = r_\square j_i(x) + \sum_{n=1}^{3} \int_{(x_n)} G_E(x-x')j_n(x')dx' \quad , \qquad (26)$$

where $i=1$, 2, or 3 are values of $n$. $E_i$ is a constant, that is individual for each strip.

The solution of this system of integral equations yields a matrix of linear inductive impedances for the strips $Z_{ij}$. To obtain the complete solution, the system must be solved 3 times, assuming that each time $j$ ($j=1$, 2, or 3) the left-hand side of the system $E_i$ takes values as follows. We assume that there is a unit linear voltage ($E_{ij} =1$ V/m) in one of the strips ($i=1$, 2, or 3) and no voltage on the two other strips ($E_{lj} =0$, $l=1$, 2, or 3 and $l \neq i$).

Each time $j$ the system is solved, we obtain three distributions of microwave current density $j_{j,i}(x)$ over cross-sections of individual strips $i$ that satisfy three conditions ($j = 1$, 2, or 3) indicated above. By integrating these current densities over the widths of individual strips, we obtain the respective total currents in the strips

$$I_{j,i} = \int_{(x_i)} j_{j,i}(x)\,dx \quad . \tag{27}$$

Ultimately, this process yields a 3x3 matrix $\mathbf{I}$ having elements $I_{ji}$. The voltage $dV_{ij}$ across a section of the length $dz$ of the strip $i$ is given by

$$dV_{ij} = E_{ij}dz. \tag{28}$$

Given our assumption $E_{ij} =1$ V/m, $E_{lj} =0$, this yields a simple expression for the matrix $\mathbf{Z}$ of the linear inductive impedances

$$\mathbf{Z} = (\mathbf{I}^T)^{-1}, \tag{29}$$

where $\mathbf{I}^T$ is the matrix transpose of $\mathbf{I}$. The diagonal elements $Z_{ii}$ of this matrix scale as self-inductances $L_{ii}$ of individual strips ($Z_{ii} = i\omega L_{ii}$), and the off-diagonal elements scale as mutual inductances $L_{ij}$ of the strips ($Z_{ij} = i\omega L_{ij}$). The inductances are complex-valued because of the inclusion of the ohmic losses in Eq. (26). Importantly, within the spin-wave band, the inductances should take complex values even in the absence of the ohmic losses in the antenna ($r_\square = 0$) due to the loss of energy of the microwave current to spin-wave excitation. In addition, one has to expect that the matrix $\mathbf{Z}$ is not symmetric within the spin-wave band, because of the non-reciprocity of magnetostatic-surface-wave (MSSW) excitation by stripline antennas [37,46]. Indeed, within the band, on top of the trivial inductive coupling of microwave currents in the strips, there is also strip coupling by propagating spin waves. Each strip of CPL excites its own spin waves in both directions. These partial waves excited by individual strips interfere to create two "total" waves that leave the CPL area, one in the positive and one in the negative direction of the axis $x$. If a partial wave, excited by an antenna strip reaches another strip of the same antenna on its way away from the antenna, it may induce an electromotive force in the strip. This effect will also contribute to the inductive coupling of the strips. Amplitudes of partial spin waves excited by a strip in the rightward and leftward directions are different. This results in different coupling strengths of that strip to its left-hand and right-hand neighbors and, ultimately, in the absence of symmetry of the matrix in the general case.

## 2.3. The input impedance of the input antenna and $S_{11}$

Once the matrix of the linear impedances has been obtained, we can solve the problem of calculating the input parameters of the input antenna. While doing so, it is important to consider the lack of symmetry of the matrix **Z**. The lack of symmetry leads to an equivalent circuit for the CPL spin-wave antenna shown in Figure 2.

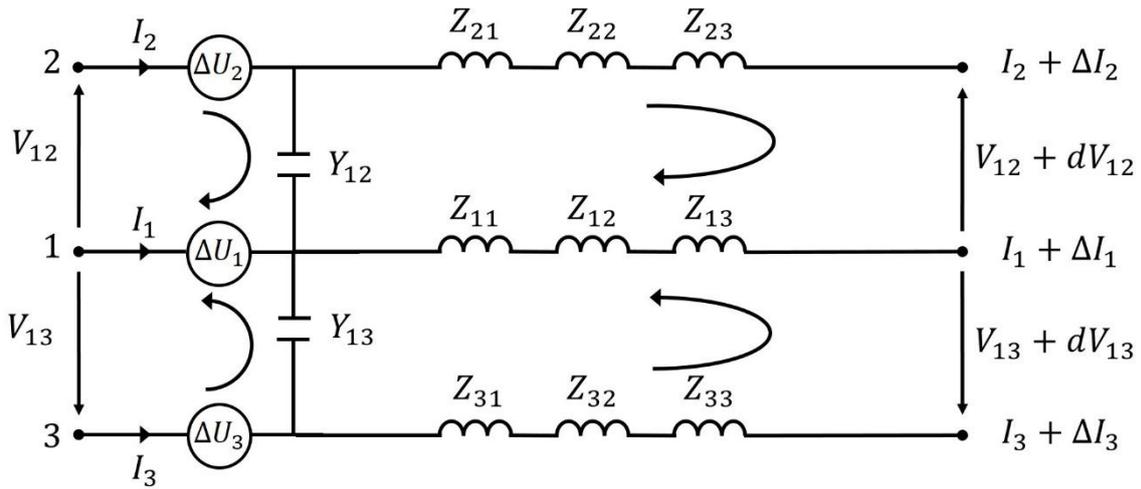

*Figure 2: An infinitesimally short segment of the CPL antenna modeled as a four-port elementary component. Conductor 1 corresponds to the signal line and conductors 2 and 3 correspond to the ground lines. V's correspond to voltages between strips. I's correspond to currents in the strips. Z's correspond to strip self and mutual (inter-strip) inductive impedances. Y's correspond to inter-strip capacitive admittance. Δu's correspond to electro-motive forces (e.m.f.) induced in the strips by an incident spin wave. The latter applies to the receiving CPL antenna only.*

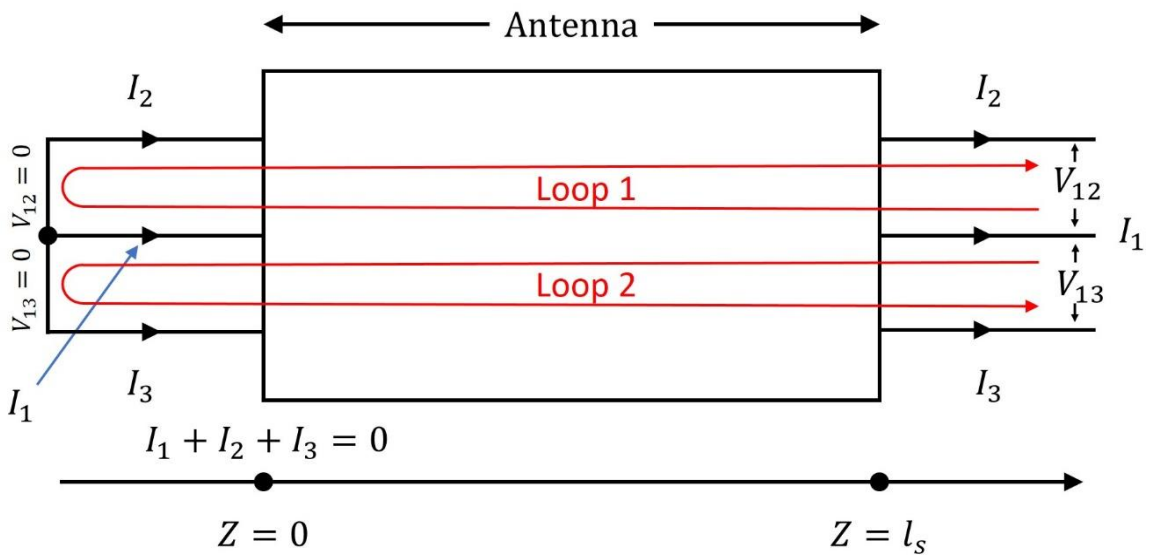

*Figure 3: Model of the CPL antenna broken up into two parallel-connected loops.*

This equivalent circuit is valid for both input and output antennas. This is because sources of voltage $\Delta U_i$ that represent electromotive forces (e.m.f.) induced in the output antenna by the incident spin wave are also included in the circuit. In addition, the equivalent circuit includes inter-strip capacitive admittances $Y_{ij}$. ($Y_{12} = Y_{13}$, and we neglect $Y_{23}$ as small with respect to $Y_{12}$ ($Y_{23}=0$).) The capacitance is due to the appearance of a microwave voltage between the strips. The voltage is due to the electric field of the electromagnetic wave propagating in the CPL. It is known that the voltage is electrostatic, therefore it is independent from the magnetostatic equations. As a result, finding the capacitance is an independent task that can be solved separately from finding a solution for the tensor $\mathbf{Z}$ (see the review paper by Vugalter [36] for more details of this idea). We calculate the capacitance self-consistently as well. The electrostatic approximation involves charges and voltages only and no currents. Therefore, the finite conductivities of the layers of the tri-layer structure do not enter the capacitance calculation, and the surface of the tri-layer metallic film can be modeled as the surface of the CPL's backing plate made of an ideal metal. The validity of such an approach can be confirmed by using the equivalent circuit of Ref. [62], which indicates a very good screening of the electric field for the parameters considered here. The self-consistent solution to this electrostatic problem is shown in Section 9 of Ref. [25].

The next step is converting the equivalent circuit from Figure 2 into a system of Telegrapher Equations for the strips. Deriving it is a straightforward procedure. One obtains a system of three equations for currents in the strips $I_i$ and two equations for inter-strip voltages $V_{12}$ and $V_{13}$ (see Figure 2). The Telegrapher Equations read:

$$\begin{aligned}
dI_1/dz &= -Y_{12}(V_{13} + V_{12}) \\
dI_2/dz &= Y_{12}V_{12} \\
dI_3/dz &= Y_{12}V_{13} \\
dV_{12}/dz &= -(Z_{11} - Z_{21})I_1 + (Z_{22} - Z_{12})I_2 + (Z_{23} - Z_{13})I_3 + \Delta U_2 - \Delta U_1 \\
dV_{13}/dz &= -(Z_{11} - Z_{31})I_1 + (Z_{32} - Z_{12})I_2 + (Z_{33} - Z_{13})I_3 + \Delta U_3 - \Delta U_1
\end{aligned} \qquad (30)$$

We cast this system of first-order differential equations into a vector-matrix form

$$d|b>/dz - \hat{A}|b> = |F>, \qquad (31)$$

and solve the vector-matrix equation for appropriate boundary conditions. This solution is placed into Section 3 of Ref. [25]. The solution yields partial input impedances, $Z_{in\,j}(j=1,2)$ of the current loops Loop1 and Loop2 of the circuit shown in Figure 3,

$$Z_{in1} = V_{12}(z=l_s)/I_2(z=l_s) = \frac{\sum_{j=1}^{5} X_{4,j} \exp(\gamma_j l_s) b_j^0}{\sum_{j=1}^{5} X_{2,j} \exp(\gamma_j l_s) b_j^0}, \qquad (32)$$

and

$$Z_{in2} = V_{13}(z = l_s)/I_3(z = l_s) = \frac{\sum_{j=1}^{5} X_{5,j} \exp(\gamma_j l_s) b_j^0}{\sum_{j=1}^{5} X_{3,j} \exp(\gamma_j l_s) b_j^0}. \tag{33}$$

where $X_{ij}$ are components of a matrix composed of the eigenvectors of $\hat{A}$, and $\gamma_j$ are the respective eigenvalues of $\hat{A}$.

It is easy to see that the vector of solution of (Eqs. (S10) of Ref. [25]) $|b_0\rangle$ scales as the microwave voltage $V_0$ incident onto the port of the input antenna. As a result, both voltages and currents that enter Eq. (32) and Eq. (33) scale as $V_0$. Therefore, the partial impedances do not depend on $V_0$, as expected. Given that, to use Eqs. (S8)-(S9) of Ref. [25] in numerical calculations, we can set $V_0$ to any constant value, for instance, $V_0=1$.

From the same diagram, Figure 4, we can also obtain an expression for the "total" input impedance for the input antenna.

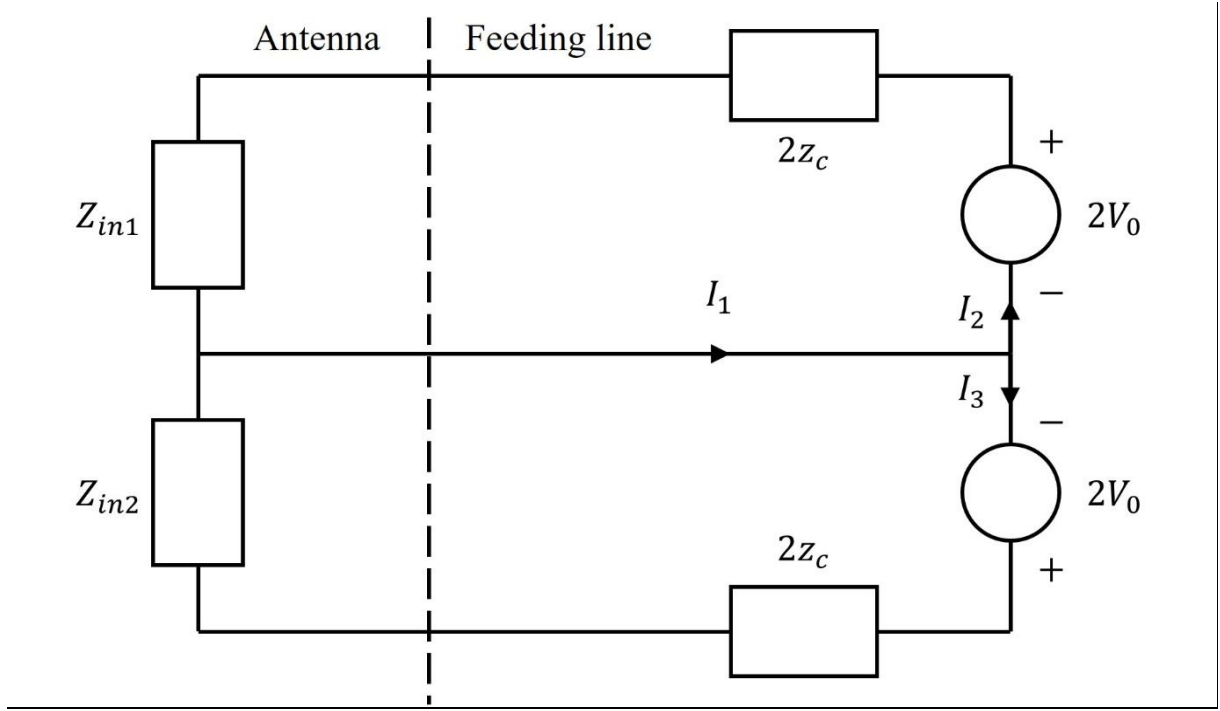

*Figure 4: Equivalent circuit of the CPL antenna connected to a feeding transmission line with characteristic impedance of $z_c$.*

From Figure 4, one sees that the two partial input impedances represent two loads connected in parallel. As a result, the total input impedance of the antenna reads

$$Z_{in\,tot} = (Z_{in1} Z_{in2})/(Z_{in1} + Z_{in2}). \tag{34}$$

The total input impedance is easily recalculated into $S_{11}$ using the standard formulas from the theory of microwave transmission lines

$$S_{11} = (Z_{int\,tot} - z_c)/(Z_{in\,tot} + z_c). \tag{35}$$

### 2.4. The model for the output antenna

The same Green's function (Eq. (22)) can be employed to model the output antenna. The first step of the modeling is the calculation of the microwave current density in the input antenna while driven by a given microwave voltage incident onto the input antenna port. We assume that the incident voltage is one volt. The equivalent circuit from Figure 4 allows for the conversion of this voltage into currents flowing through the strips of the antenna.

The input current of the signal line (Strip 1) is given by

$$I_{in1} = 1Volt \left[ \frac{2}{Z_{in1} + 2z_c} + \frac{2}{(Z_{in2} + 2z_c)} \right] = \frac{2}{(Z_{in1} + 2z_c)} + \frac{2}{(Z_{in2} + 2z_c)}, \tag{36}$$

and the input currents of the ground lines (Strips 2 and 3 respectively) by

$$I_{in2(3)} = 1Volt[2/(Z_{in2(3)} + 2z_c)] = 2/(Z_{in2(3)} + 2z_c). \tag{37}$$

In Section 4 of Ref. [25], we show that the current in the antenna strips for the microscopic geometries of interest $(\gamma_j l_s \ll 1)$ is practically uniform along the antenna. Then we can assume that they are equal to the input currents:

$$I_i(z) \cong I_{in\,i}. \tag{38}$$

The electric field $\mathbf{E}(z)$ of the excited spin waves at an arbitrary position along the input antenna then reads

$$\mathbf{E}(z) = \mathbf{Z} \cdot \mathbf{I}(z). \tag{39}$$

Because of our assumption in Eq. (38), Eq. (39) reduces to

$$\mathbf{E}(z) \cong \mathbf{Z} \cdot \mathbf{I}_{in}, \tag{40}$$

where $\mathbf{I}(z)$ and $\mathbf{I}_{in}$ are column-vectors containing entries $I_i(z)$ and $I_{in\,i}$ respectively, and $\mathbf{E}(z)$ is a vector containing the electric fields $E_i(z)$ in strips $i=1,2,3$ as its components. Eq. (40) implies that the electric fields are constant along the strip lengths $E_i(z)= E_i$. Thus, for the short antennas, the electric field is uniform over the antenna length.

If we now substitute these $E_i$ values into the integral equation (Eq. (26)) as its left-hand side and solve the equation with respect to $j_n(x)$ we obtain the current density distribution for 1 Volt of microwave voltage incident onto the antenna (see Eqs. (36)-(37)).

The last step is to substitute the obtained current density into Eq. (23) and treat the left-hand side of the expression as an unknown. This yields an expression for the electric field of the spin wave at any position $x$ in the film, including beneath the output antenna

$$E^{SW}(x) = \sum_{n=1}^{3} \int_{(x_n)} G_E(x-x') j_n(x') dx' \quad , \tag{41}$$

where $G_E(x-x')$ is given by Eq. (22). Recall that the three integrals involving $G_E(x-x')$ are taken over the widths of the respective strips $n=1, 2,$ and 3 of the input antenna.

Once we have obtained the expression (Eq. (41)) for the electric field of the spin wave at any location $x$ in the film, we can proceed with the calculation of the output parameters of the output antenna. For the output antenna, we do not have a source of external microwave voltage. Instead, we have voltage sources $\Delta U_i$ in the antenna (Figure 2). The voltage sources originate from the electric field of the spin wave incident onto the output antenna (Eq. (41)). However, they are not the same as the field of the incident spin wave [36]. The electric field of the wave has a form of a traveling wave. Therefore, it is non-uniform over the widths of the antenna strips. However, $\Delta U_i$ must be uniform across the width of a metallic strip carrying a microwave current. The reason for this requirement is the same as for the requirement of the uniformity of the linear voltage induced in the strips of the input antenna by the dynamic magnetization in the ferromagnetic layer (see Eq. (25)). This observation tells us that the electric field of the incident spin wave must generate not only global (or signal) currents in the strips, that obey the Telegrapher Equations, but also local currents $j_n^{(l)}$ that re-distribute electric charges across the width of each strip. This gives rise to width-uniform e.m.f. $\Delta U_i$.

The local currents must not contribute to the currents that obey the Telegrapher Equations. This requirement gives rise to an integral equation as shown in Section 11 of Ref. [25]. Solving this equation yields a simple formula for numerically calculating $\Delta U_i$. This formula reads

$$\Delta U_i = \sum_{j=1}^{3} Z_{ij} I_{inc\ j} \quad , \tag{42}$$

where $Z_{ij}$ is the respective component of the matrix **Z** Eq. (29), and the three components $I_{inc\ i}$ of the vector $\mathbf{I}_{inc\ i}$ are obtained by solving the system of integral equations that has the same kernel as Eq. (26)

$$r_\square j_{inc\ i}(x) + \sum_{j=1}^{3} \int_{(x_j)} G_E(x-x') j_{inc\ j}(x') dx' = E_{mc\ i}(x) \quad . \tag{43}$$

This system is the CPL-analogue of an integral equation (Eq. (S66)) from Section 10 of Ref. [25]. The variables $x$ and $x'$ take only values within the coordinate ranges $(x_i)$ for the antenna strips $i=1,2,3$. Accordingly, $E_{mc\,i}(x)$ is the same electric field of the incident spin wave as $E^{SW}(x)$ from Eq. (41) but taken only within the co-ordinate range $(x_i)$ for the $i$-th strip of the antenna. Similarly, the functions $j_{inc\,i}(x)$ are defined within the ranges $(x_i)$ only. Once the three functions $j_{inc\,i}(x)$ have been found, the components of the vector $\mathbf{I}_{inc}$ are obtained by integrating the functions over the widths of the respective antenna strips

$$I_{inc\,i} = \int_{(x_i)} j_{inc\,i}(x)dx \ . \qquad (44)$$

This reduces the modeling of the output voltage for the output antenna to solving the same Telegrapher Equations (Eqs. (30)) but now with the three obtained e.m.f. $\Delta U_i$, $i=1, 2, 3$ (Eq. (42)) included into the equations. They represent distributed sources of microwave electric voltage (see Figure 2).

As already found, the matrix of linear inductances and capacitances $\hat{A}$ (Eq. (31) for these equations is the same as for the input CPL. Physically, this is because the process is the same – there is some primary voltage (an external source $V_0$ for the input CPL (Figure 4), or "built-in" sources $\Delta U_i$ for the receiving one) that generate global (or signal) currents in the strips. The presence of the primary voltage ultimately yields a secondary voltage that scales linearly with the signal current. The scaling tensor is the same matrix $\hat{A}$ of inductances/capacitances/resistances. Note that from this point of view, the process of the reception of a signal carried by a spin wave can be considered as the scattering of the incident spin wave from the system of the three parallel metal strips of the output CPL. This is explained in Section 11 of Ref. [25].

Solving Eqs. (30) for the output antenna yields $S_{21}$. The solution is given by the complete expression shown in Eq. (S4) in Ref. [25] i.e. with the non-vanishing vector $|F>$. The solution must be complemented by appropriate boundary conditions. The boundary conditions that we assume are the connection of the output port of the output antenna to a section of a coplanar transmission line. This is in agreement with the conditions of the experiment in [22] – a section of the microscopic coplanar line picks up the output-antenna signal and carries it to the CPL contact pads of a microscopic microwave probe. Section 5 of Ref. [25] implements this idea and converts the solution given by Eq. (S4) into the output voltage $V_{out}$ of the output port of the device.

The complex-valued $S_{21}$ is then obtained by dividing $V_{out}$ by the microwave voltage incident from the generator onto the input port of the input antenna. Recall that we assumed that this voltage amounts to 1 Volt (see Eqs. (36)-(37)). Therefore, $S_{21}$ is just $V_{out}$.

The efficiency of device transmission in the opposite direction – $S_{12}$ – can then be calculated either by simultaneously changing the signs of the applied field and static magnetization to the opposite ones in the final formulas, or by placing the receiving antenna to the left ($x_i<0$) of the input antenna.

Note that most earlier theories relied on an assumption that the efficiency of reception of spin waves by an antenna is the same as of excitation of spin waves by it. This idea is a consequence of the Lorentz Reciprocity Theorem [63]. In Ref. [64] this was rigorously shown for a microstrip antenna. Therefore, it was important to check that the present theory agrees with this antenna property. We did this check numerically and found that our equations satisfy this physical requirement with very high numerical precision.

## 3. Modeling results

### 3.1. Numerical implementation of the theory

Because there is no analytical solution for the Green's function of the electric field (Eq. (22)), the only way to solve the integral equation (Eq. (26)) is by using numerics. We solve the integral equation by converting the integral into a discrete sum over a linear mesh that covers the strips of the coplanar line. There are no mesh points in the gaps between the strips because the current density in the gaps is zero. We divide the cross-section (in the direction $x$) into 100 equal intervals $\Delta x$. This converts the integral equation into a system of 100 linear algebraic equations

$$E_i = r_\square j_i(x_{j'}) + \Delta x \sum_{j=1}^{100} E_z^0(x_{j'} - x_j) j_n(x_j) . \qquad (45)$$

The unknowns of the system are values of current density $j_i(x_j)$ at the mesh points $x_j$, $j=1,2,..100$. The inhomogeneous terms of the equations are values of the electric field $E_i$ in the strips. (Recall the field is uniform across individual strips $i$.)

The coefficients $E_z^0(x_{j'} - x_j)$ of the system of equations are values of the Green's function (Eq. (22)) calculated for a pair of points $x_{j'}, x_j$. Because the Green's function is translationally invariant, its values depend on the distance $s_{j,j'}$ between the points, but not on the absolute positions of the points. Therefore, from 100x100 values of the coefficients of Eq. (45), only 200 are different. (Due to the non-reciprocity of MSSW excitation by stripline antennas, we have $E_z^0(s_{j'j}) \neq E_z^0(s_{jj'})$, therefore the 200=2x100 values). Calculating the 200 values of the Green's function represents the most time-consuming step of the simulation because the analytical solution for the Green's function is obtained in the Fourier space (see Eq. (S31) in Ref. [25]). The inverse Fourier transform of Eq. (S31) is given by Eq. (22). The integral in Eq. (22) is evaluated numerically using the adaptive integration method. Because of the complicated structure of the terms entering Eq. (S31), the calculation of the Fourier integral Eq. (22) for each separate value of $s_{j,j'}$ takes noticeable computer time. As a result, building the array of $E_z^0(s_{j'j})$ takes most of the time of the whole simulation process. Once the Green's function values have been calculated, and the matrix of the coefficients of Eq. (45) has been constructed based on those values, Eq. (45) is solved using numerical methods of linear algebra.

We use the same method of adaptive integration to evaluate the integral of Eq. (41). To this end, we divide the cross-section of the output coplanar antenna into the same 100 intervals, as the input antenna. Similar to the treatment of the input antenna, we use the translational invariance of the Green's function to reduce the number of times it needs to be calculated.

The numerical calculation of the linear capacitance of the coplanar line (see Section 9 of Ref. [25]) proceeds in a similar way. The only difference is that because the capacitance is calculated in the electrostatic approximation, it needs to be calculated just once, whereas we need to go through the whole process again each time we change the frequency of the input microwave signal while calculating the **Z** matrix (Eq. (29)).

Once the values of the components of the matrix **Z** have been computed, the remainder of the simulation is straightforward. The next step is to convert the matrix into the matrix $\hat{A}$ (Eq. (S3) in Ref. [25]) and calculate the eigenvalues and eigenvectors of $\hat{A}$, using numerical methods of Linear Algebra. This result is used to compute the vector $|b_0\rangle$ (Eq. (S4) of Ref. [25]), then the partial input impedances (Eqs. (32)-(33)) and the total input impedance of the input antenna from them (Eq. (34)). The input impedance is then converted into the reflection coefficient $S_{11}$ by using the standard formula from Eq. (35), assuming the characteristic impedance of the feeding line of 50 Ohm. This concludes the computation of $S_{11}$.

In order to calculate $S_{21}$, we first calculate the currents in the input antenna strips, assuming that a microwave voltage of 1 Volt is incident onto it from the feeding line. We use the expressions in Eqs. (36)-(38) and the same matrix **Z** as for computing $S_{11}$ to this end. Equation (40) is then used to calculate the linear voltages in the antenna strips that correspond to these currents. The next step is to numerically solve the integral equation (Eq. (26)) to compute the respective current densities, employing the obtained linear voltages as its left-hand side. Then the found current densities are substituted into Eq. (41) in order to compute a set of values of the electric field $E^{sw}(x_i)$ of the excited SW for ranges of coordinates $x_i$ that correspond to the strips of the output antenna.

Once $E^{sw}(x_i)$ has been obtained, we substitute them into the right-hand side of Eq. (43) and solve that integral equation for $j_{inc\ i}(x_i)$. We apply the same method of converting it into a matrix-vector equation as employed for computing the components of **Z** (Eq. *(29))*, but now the co-ordinate mesh spans the output antenna. Note that the matrix of the equation is the same as for computing **Z** because of the translation invariance of the Green's function of the electric field of the spin wave. Therefore, the already-computed matrix of Eq. *(29)* can be recycled, which accelerates the solution. Then we employ Eqs. (44) and (42), to compute the e.m.f. values $\Delta U_i$. The final step is to compute $V_{12}$ and $V_{13}$ using Eq. (S13) of Ref. [25] and employing the obtained e.m.f. values. They enter the vector $|F\rangle$ of Eq. (S14) of Ref. [25], which is defined by Eq. (S2) of Ref. [25]. The calculation concludes by converting $V_{12}$ and $V_{13}$ into a single value $V_{out}$ using Eq. (S17) of Ref. [25]. Given that we assumed 1 Volt of voltage incident onto the input antenna, we have $S_{21}= V_{out}/(1\ \text{Volt})= V_{out}$.

The numerical code that implements this algorithm is available from GitHub [https://github.com/KingHarley/SpinWave-Simulations-Public].

### 3.2. Simulation Results

In the pages that follow, about section 3.2, we employ our theoretical model to investigate and discuss some of the details associated with the excitation and reception of spin waves by CPL antennas. In section 3.2.1 we investigate the *S*-parameters of the simulation and check the validity of the shape of their amplitude curves, the frequency position of the maxima of the amplitude curves, and the group velocity of the spin waves, by means of looking at the phase of the *S*-parameters, against an external model of the dispersion relation. In section 3.2.2 we investigate the presence of a frequency non-reciprocity in the simulation data even for the case of a fully symmetric system. In section 3.2.3 we discuss the input impedance of the CPL antenna which underlies the formation of the scattering parameters. In section 3.2.4 we show the electric field associated with the spin-wave signal and show its amplitude as a function of the *x*-distance in the ferromagnetic stripe. It is this electric field that is detected by the reception antenna. In section 3.2.5 we investigate the effects of the material parameters and how they affect the shape, frequency position, and phase of the *S*-parameters when simulating the spin-wave signal. In particular, we investigate the effects of the perpendicular magnetic anisotropy, interfacial Dzyaloshinskii-Moriya interaction, saturation magnetization, gyromagnetic ratio, Gilbert damping constant, and the externally applied magnetic field. The latter results are placed into the online supplemental materials [25] in order to shorten the paper.

The model was implemented as MathCAD and Python codes. The Python code appeared to be much faster, and below we present the results of simulations made with this software.

In the example calculations and graphs that follow we use a default set of system parameters shown in Table 1. Assume that all calculations were done for the default system parameters unless explicitly mentioned.

*Table 1: Default system parameter values. w, w$_g$, and Δ$_g$ are the widths of the signal line, ground lines, and gaps between signal and ground lines, of the coplanar antenna, respectively. l$_s$ is the length of the coplanar antenna. l$_d$ is the separation distance of the exciting and receiving antennas, measured from their centers. 4πM$_s$ is the saturation magnetization of the ferromagnetic layer. γ is the gyromagnetic ratio. H is the applied external field. α$_G$ is the Gilbert damping constant. A is the exchange constant. L is the thickness of the ferromagnetic material. d$_1$ and d$_2$ are the thicknesses of the non-magnetic layers interfaced to the ferromagnetic layer, and d$_s$ is the thickness of the silicon-oxide layer separating the top magnetic layer from the coplanar antennas. K$_{u1}$, K$_{u2}$, D$_1$, and D$_2$ are the surface PMA constants and surface DMI constants at the interfaces with non-magnetic metals of thicknesses d$_1$ and d$_2$ respectively. σ, σ$_1$, and σ$_2$ are the bulk conductivities of the ferromagnetic metal, and the bottom (d$_1$) and top (d$_2$) non-magnetic metals. ε$_s$ is the dielectric constant of the SiO$_2$ spacer layer. ρ$_{Al}$ is the resistivity of the aluminum in the coplanar antennas.*

| **w (nm)** | 648 | **d$_1$ (nm)** | 5 | **$\sigma_2$ (10$^7$ S/m)** | 0.952 |
|---|---|---|---|---|---|
| **w$_g$ (nm)** | 324 | **d$_2$ (nm)** | 5 | **$\varepsilon_s$** | 3.8 |
| **Δ$_g$ (nm)** | 334 | **d$_s$ (nm)** | 130 | **$\rho_{Al}$ (10$^{-8}$ Ω m)** | 2.65 |
| **l$_s$ (μm)** | 20 | **4πM$_s$ (G)** | 20900 | **$K_{u1}$ (mJ/m$^2$)** | 0 |
| **l$_d$ (μm)** | 2.464 | **γ (MHz/Oe)** | 3 | **$K_{u2}$ (mJ/m$^2$)** | 0 |
| **$\alpha_G$** | 0.0107 | **H (Oe)** | 1000 | **$D_1$ (pJ/m)** | 0 |
| **A (10$^{-11}$ J/m)** | 2.625 | **σ (10$^7$ S/m)** | 1.7 | **$D_2$ (pJ/m)** | 0 |
| **L (nm)** | 20 | **$\sigma_1$ (10$^7$ S/m)** | 1.409 | | |

### 3.2.1. Scattering parameters ($S_{11}$, $S_{21}$, and $S_{12}$)

The simulated *S*-parameters are shown in Figure 5. The reflection, $S_{11}$, parameter is given by Eq. (35) and is identical to the $S_{22}$ parameter due to the symmetry of the CPL antennas in the simulation. The $S_{21}$ parameter is given by Eq. (S17) and the $S_{12}$ parameter is calculated by placing the receiving antenna to the left ($x_i<0$) of the input antenna.

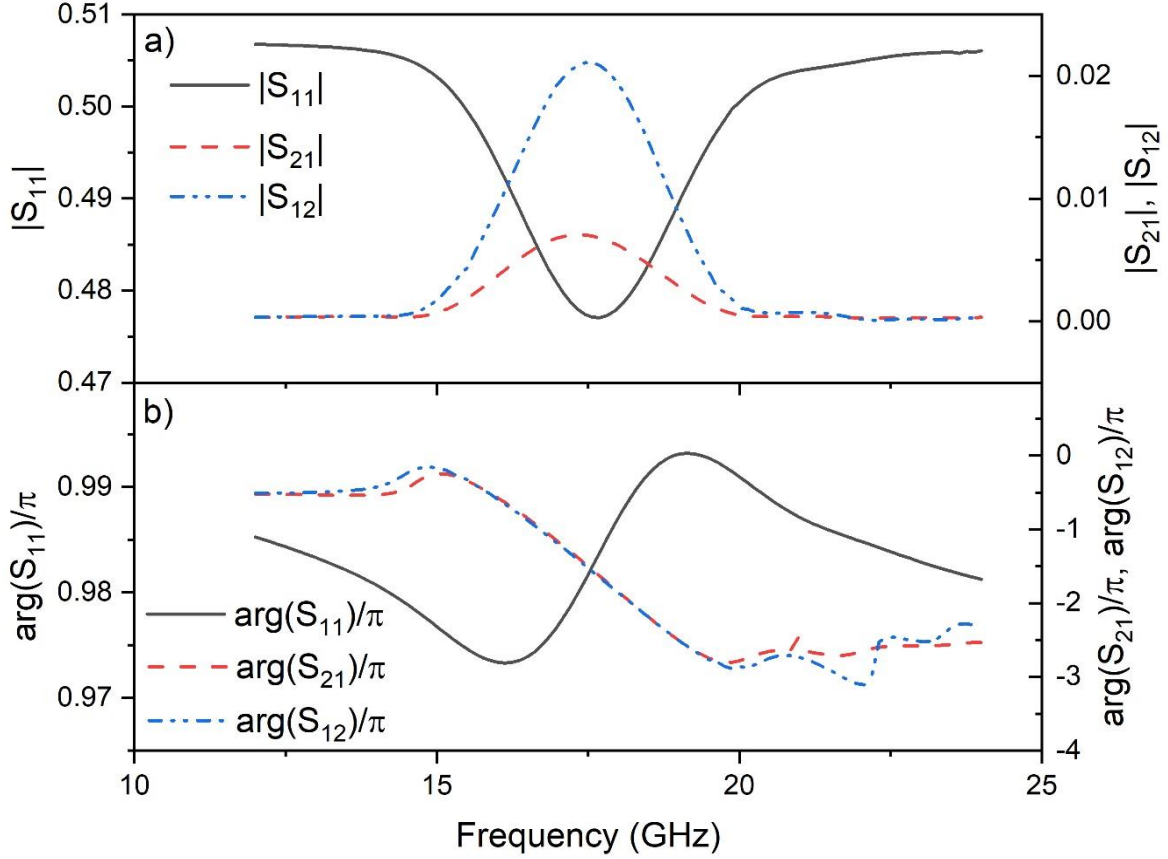

*Figure 5: a) Amplitude of the S-parameters. b) Phase of the S-parameters scaled by 1/π. Black-solid line: $S_{11}$, red-dashed line: $S_{21}$, blue-dash-dotted line: $S_{12}$.*

As the characteristic impedance of the feeding lines does not change, only the load impedance of the antenna affects the $S_{11}$ coefficient. During the spin-wave excitation, the input impedance (Eq. (34)) changes as a result of radiation and damping losses which alters the $S_{11}$ parameter. The amplitude of the $S_{11}$ parameter resembles a Lorentzian line shape, and the minimum of reflection occurs when the spin-wave excitation is maximized. From Figure 5(a) the minimum reflection occurs at ~18.054 GHz. The phase of the $S_{11}$ characteristic sits around π radians which evidences that the magnitude of antenna input impedance is smaller than the characteristic impedance of the feeding line.

This is expected. Off of the spin-wave band, the antenna represents a section of a coplanar line, short-ended at its far end, which is much shorter than the wavelength of microwaves in the antenna. In these circumstances, the input impedance is small and is mostly due to ohmic losses in the antenna. On the spin-wave band, power losses due to the conversion of input power into spin-wave power add to the ohmic losses. Because of the very small length of the antenna and small thickness of the ferromagnetic layer, it is impossible to convert a significant part of the incident microwave power into the power of traveling spin waves. Therefore, the input impedance remains small on the band, resulting in the phase of the coefficient of reflection, of the incident microwave signal from the input port of the antenna, to remain close to $\pi$.

The shapes of the transmission characteristics, $S_{21}$ and $S_{12}$, follow from the spin-wave band which is excited by the input antenna and shown by the $S_{11}$ parameter. The maximum of the transmission characteristics correspond to the maximum spin-wave excitation which align well with the minimum of reflection. This is expected since the spin-wave excitation acts as a way of energy absorption by the system and thus will reduce the reflection of the system. The maximum transmission for $S_{21}$ and $S_{12}$ occurs at ~17.886 GHz and ~17.982 GHz, respectively. Note that the two values are not equal to each other which is an interesting finding in this simulation. It will be addressed later on.

Outside of the spin-wave band, the phase of the transmission $S_{21}$ and $S_{12}$ is governed entirely by the direct inductive coupling of the input antenna with the output antenna and as a result, remains almost constant. Inside the spin-wave band, the phase becomes non-constant as a result of the phase accumulation of the propagating spin wave. This can be seen as the almost linear regime of the phase profile between ~16 and 20 GHz in Figure 5(b). If we ignore a potential contribution of the processes of reception and emission of spin waves by the antennas to the transmission phase, then the slope of the phase is governed by the slope of the dispersion relation, also known as the group velocity of the MSSW. For the propagating waves, the phase is given by:

$$\varphi = kx, \qquad (46)$$

where $k$ is the wavenumber of the wave and $x$ is the distance the wave has traveled. As mentioned, the group velocity is given by the slope of the dispersion relation,

$$v_g = \frac{\partial \omega}{\partial k} = \frac{x}{\Delta_\varphi}, \qquad (47)$$

where $\Delta_\varphi = \frac{\partial \varphi}{\partial \omega}$ is the slope of the phase profile of the MSSW (for a distance of propagation $x$) which is the same as the transmission characteristic. Thus if one knows the propagation distance $x$ of the spin wave then it would be possible to calculate the group velocity of the spin wave using the slope of the phase given in Figure 5(b). However, determining the propagating distance $x$ is not straightforward as the antennas have finite widths in the direction $x$, and it is not clear from which points on the two spin-wave antennas one should measure the propagation distance. However, given the propagation distance remains constant, one may use the slope of the phase to gain an understanding of how the group velocity behaves. From Figure 5(b) it is clear that for the default system parameters the forward and backward propagating waves have slightly different group velocities, with the wave traveling in the –$x$-direction ($S_{12}$) having a larger group velocity. This is seen from the smaller slope of the phase of $S_{12}$ in the spin-wave band. This is investigated further in the next section.

Let us compare the wavenumbers obtained from the dispersion relation (Figure S2) to Eq. (S34) (both from [25]) which describes the electric field of a spin wave in the Fourier space $e_{zkm}(y_0)$, measured at the height of the antenna $y_0$ and excited by an infinitely thin wire. The wavenumber dependence of the amplitude of the electric field follows from the solution of the boundary-value problem Eqs. (17)-(19). The solution of the linear system of equations (Eqs. (17)-(19)) is maximized when the determinant of the system is minimized. The maximum of the solution corresponds to the maximum of $e_{zkm}$ as a function of $k$. Our numerical calculations show that the maximum is sharp (see Figure 6). We expect its position $k$ for a given excitation frequency $\omega$ to agree with the numerical spin-wave dispersion model (Figure S2) because a vanishing of the determinant represents an alternative formulation of the spin-wave dispersion law.

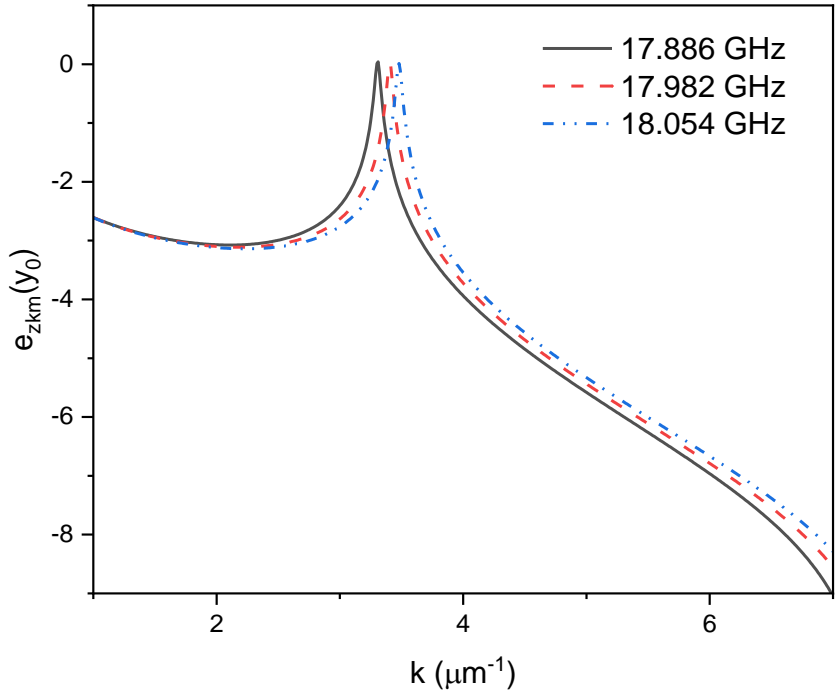

*Figure 6: Numerical calculation of $e_{zkm}(y_0)$ as a function of the Fourier wavenumber k for 17.886 GHz (black-solid line), 17.982 GHz (red-dashed line), and 18.054 GHz (blue-dash-dotted line). The maxima for 17.886, 17.982, and 18.054 GHz correspond to 3.31, 3.40, and 3.48 $\mu m^{-1}$, respectively.*

For the frequencies, 17.886, 17.982, and 18.054 GHz, Eq. (S34), of Ref. [25], observes a maximum at 3.31, 3.40, and 3.48 $\mu m^{-1}$, respectively. Note that only positive wavenumbers are checked here since the dispersion relation is symmetric around $k$=0 for the default system parameters. These values were calculated for the case of no magnetic losses in the system ($\alpha_G = 0$) since the numerical dispersion model used above does not include magnetic losses either (but it still includes ohmic losses due to finite conductivities). Both methods give identical results which confirms that the underlying physics of the CPL theory is consistent with an independent numerical model for the dispersion relation (see Section 6 of Ref. [25]).

### 3.2.2. Frequency non-reciprocity for a symmetric system

It is clear from Figure 5 that there is an amplitude and frequency non-reciprocity between the forward and backward traveling waves. The magnetostatic surface spin wave (MSSW) considered in this work, is characterized by non-reciprocity of its wave properties [65]. This means that two opposite directions of the wave propagation (+*x* and –*x*) are not equivalent. In particular, the wave is excited more efficiently in one direction than in the opposite one, and the direction of the more efficient excitation depends on the direction of the applied field. In addition, the MSSWs localize at one of the surfaces of the ferromagnetic layer depending on the propagation direction. It is important to keep in mind that the wave localization at a particular interface of a ferromagnetic layer is not the main reason for the non-reciprocity of MSSW excitation by stripline antennas. This was originally shown in Ref. [31]. Ref. [66] explains this more transparently. The main contribution to the non-reciprocity is the interplay of spin-wave polarization with the polarization of the microwave magnetic field of the antenna. The wave localization at one of the interfaces of the ferromagnetic layer does not contribute significantly to the process. Furthermore, as shown in Ref. [67] and experimentally confirmed in [68], for very thin ferromagnetic layers, for which the effect of the inhomogeneous exchange interaction on the magnetization dynamics is very significant, the wave is localized at the interface opposite to the interface of localization of the usual Damon-Eshbach dipole-dominated MSSW. Furthermore, the localization is quite weak – for *k*=3.45 $\mu m^{-1}$, which is within the typical range of MSSW excitation by a microscopic CPL antenna (see Figure 5), the difference in $m_x$ amplitudes for a 20nm thick cobalt layer with neither PMA nor iDMI at its interfaces is just 1%. This result is obtained with the numerical code from Ref. [67]. This almost vanishing localization cannot noticeably contribute to the non-reciprocity of excitation.

The frequency non-reciprocity between the maxima of the $S_{21}$ and $S_{12}$ characteristics is ~96 MHz and is not understood for the default system parameters chosen. It is well known that the presence of further symmetry breaking through the inclusion of PMA or DMI will introduce a frequency non-reciprocity in the dispersion relation of the spin waves. However, for the chosen parameters, which resulted in Figure 5, there was no inclusion of PMA or DMI. Thus the only sources of asymmetry in the system are the localization of the CPL antenna at one surface, the direction of the applied magnetic field, the inclusion of the two non-magnetic metal conducting layers above and below the ferromagnetic layer, and the dielectric spacer layer separating the CPL from the metal multilayer. Whereas the localization of the CPL antenna and the direction of the applied magnetic field are fundamental to the geometry and cannot be symmetrized, one can remove the non-magnetic metal conducting layers and dielectric spacer.

To check this we set the conductivities of the non-magnetic metal layers to vanishingly small values $\sigma_1 = \sigma_2 = 1$ Ωm and the dielectric constant of the spacer $\varepsilon_s = 1$, and made the thicknesses of these layers small compared to the ferromagnetic layer ($d_1 = d_2 = d_s = 0.01$ nm). The remaining parameters are the same as the default parameters in Table 1. The resultant transmission characteristics $S_{21}$ and $S_{12}$ are shown in Figure 7.

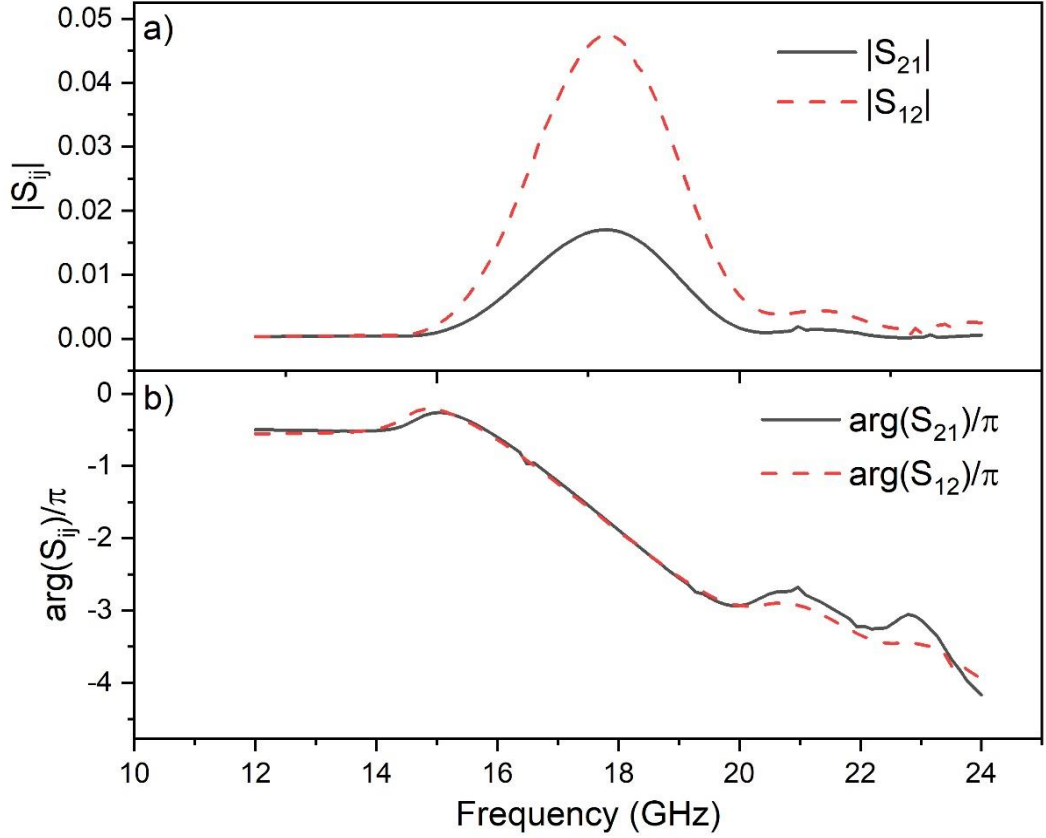

*Figure 7: Amplitude (a) and phase (b) of the transmission characteristic for the fully symmetric system. Black-solid line: $S_{21}$. Red-dashed line: $S_{12}$*

The maximum transmission for the forward ($S_{21}$) and backward ($S_{12}$) traveling waves occur at 18.150 and 18.174 GHz, respectively, which correspond to 3.58 and 3.60 $\mu m^{-1}$, respectively. Thus, for this fully symmetric system, there is a frequency non-reciprocity of 24 MHz, which is smaller than the 96 MHz obtained for the default system parameters. This suggests that the major part of the frequency non-reciprocity extracted for nominal system parameters stems from the asymmetric electromagnetic environment of the film (conducting and dielectric neighboring layers). The remaining part is non-negligible. The only explanation for this residual frequency non-reciprocity remains the polarization of the applied field and localization of the CPL antenna. This has not been observed in previous theoretical studies [26–39] and thus is a consequence of the fully self-consistent approach to the problem of the excitation of spin waves by a CPL antenna developed in this study.

Let us now look at the current distribution and its Fourier transform for the input antenna at 18.150 GHz, which corresponds to the maximum transmission of $S_{21}$ for this maximally symmetrized system (shown by the black-solid line in Figure 8). The current distribution at 18.174 GHz, corresponding to the maximum transmission of $S_{12}$, overlaps perfectly with the current distribution at 18.15 GHz and thus is not shown in Figure 8. Thus, the current distribution does not change significantly for frequency differences of < 24 MHz. We also check the current distribution in the antennas when omitting the presence of ohmic losses in the uniform electric field across the strips (Eq. (26)), when calculating the current distribution.

For this latter case, the current distribution is checked at 18.16 GHz and is shown by the red-dashed line in Figure 8. Note the strong left-right asymmetry, which is attributed to a non-reciprocal back-action of the spin-waves onto the microwave current, a phenomenon which could be of importance for thicker magnetic films and/or antennas of larger lateral size [46].

In our case, the inclusion of ohmic losses in the antenna strips acts to redistribute the current across the strips and results in a much more uniform current distribution than in the case where the ohmic losses are not included, as seen in Figure 8(a, b). This is expected since the antenna widths are smaller than the skin depth of aluminum (~700 nm) and thus one expects the current to distribute rather uniformly. The more uniform and, hence, more symmetric the current distribution is, the more symmetric its Fourier transform will be around $k=0$, as seen in Figure 8(c). The geometry of the excitation antenna confines the range of wavenumbers it can excite and the nominal wavenumber corresponds to the wavenumber which is excited most efficiently by the antenna. This wavenumber has typically been determined by the maximum of the Fourier transform of the current distribution in the excitation antenna. This nominal wavenumber does not depend on the material parameters and thus is solely a result of the spatial antenna geometry. Furthermore, as long as the dispersion relation exists for the nominal wavenumber given by the excitation antenna, it is expected that the wavenumber of maximum transmission, detected at the receiving antenna, corresponds to the same nominal wavenumber of the excitation antenna. The two main maximum peaks of the Fourier transform correspond to +3.58 and -3.63 $\mu m^{-1}$ at 18.160 GHz when including ohmic losses and to +3.17 and -3.48 $\mu m^{-1}$ at 18.16 GHz when omitting ohmic losses. The nominal wavenumbers extracted when including ohmic losses are almost identical for the two propagation directions and match up almost perfectly with the wavenumbers extracted from the dispersion relation. This reinforces the method of using the Fourier transform to determine the nominal wavenumber.

The asymmetry seen in Figure 8 is a result of the self-consistent solution. Thus, this more precise solution demonstrates that the non-reciprocity of spin-wave excitation by the input antenna may result in a "residual" non-reciprocity of spin-wave transmission not accounted for in other self-consistent theories [36,38]. This result is important because quantifying the frequency non-reciprocity of counter-propagating spin waves underlies the method of traveling-spin-wave spectroscopy for extraction of surface PMA and iDMI [14,17,22]. The utilization of meander-like antennas [14,17,21,22], and a comparison of the phase of the two signals rather than of their amplitude [24,68] should help to overcome this problem.

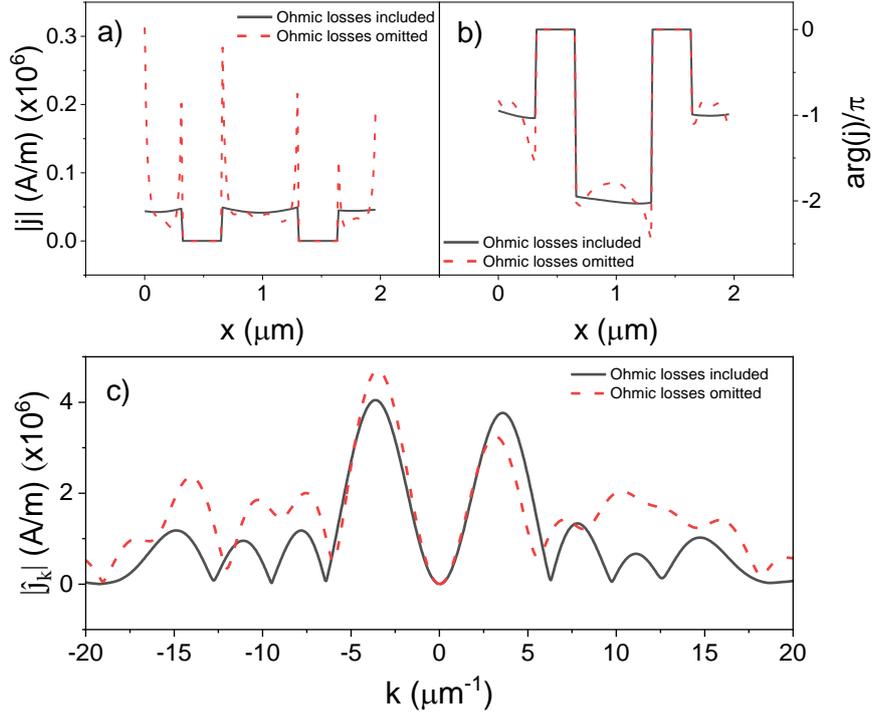

*Figure 8: The amplitude (a) and phase (b) of the current density, j, at the x position of the input antenna for the maximally symmetrized system. c) The spatial Fourier transform of the current density, j, is shown in a) and b). Black-solid line: 18.160 GHz. Red-dashed line: 18.255 GHz.*

### 3.2.3. Formation of $S_{11}$ and $S_{22}$

Underlying the $S_{11}$ parameter is the input load impedance $Z_{in\,tot}$ (Eq. (34)) which is governed by the impedance matrix **Z** (Eq. (29)) of the input antenna. Since the output antenna has the same geometry as the input antenna, the $S_{11}$ and $S_{22}$ parameters are governed by the same impedance matrix **Z**. The diagonal components of **Z** correspond to the self-inductances of the individual strips in the antennas, whilst the off-diagonal components correspond to the mutual inductances between the strips. It is known that a single wire antenna will have a broadband spectrum in the frequency space [31,34,36,37], whilst a coplanar or meandering antenna has a much narrower band spectrum in frequency space [19,21,22]. Figure 9 shows the real part of the input load impedance as well as the real part of the $Z_{11}$ component of the **Z** matrix as a function of frequency, calculated for the default system parameters as well as for the maximally symmetrized system. The $Z_{11}$ component of the **Z** matrix corresponds to the self-inductance of the signal strip, thus this value can be used as a proxy for the scalar radiation impedance of a single-strip antenna.

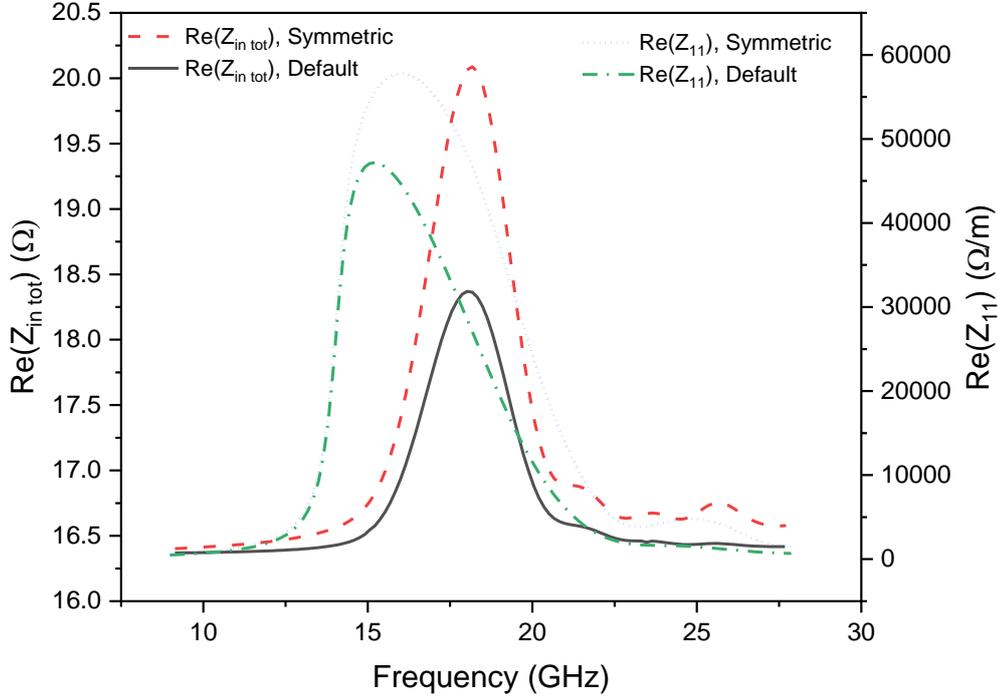

*Figure 9: Black-solid (red-dashed) line: Real part of the total input impedance, $Z_{in\,tot}$, for default (maximally symmetrized) system parameters. Green-dash-dotted (blue-dotted) line: Real part of the $Z_{11}$ component of the **Z** matrix corresponding to the self-impedance of the signal line, for default (maximally symmetrized) system parameters. (Note a linear slope was subtracted from $Re(Z_{11})$ to allow better comparison. The linear slope is due to the off-spin-wave-band self-inductance of the antenna that contributes to the real part of $Z_{11}$ through the presence of ohmic losses of eddy currents induced in the magnetic and non-magnetic layers of the layered structure by the microwave currents in the antenna.)*

From Figure 9 it is evident that the single wire $Z_{11}$ has a broader band than the CPL described by $Z_{in\,tot}$. The maximum peak of $Z_{11}$ begins to form at ~12 GHz and reaches a maximum upwards slope at ~14 GHz. For a single-strip antenna, the maximum transmission is expected to sit at $k=0$, which for the chosen system would correspond to 14.039 GHz. This aligns well with the beginning of the $Z_{11}$ peak.

From the engineering point of view, single-strip antennas are not convenient in the microscopic geometries, and CPL antennas lithographically formed on top of a layered magnetic structure are largely utilized instead. One important advantage of the CPL geometry is that the characteristic impedance of the antenna off the spin-wave band can be easily matched to the characteristic impedance of the feeding line. From Figure 9 one sees that the CPL has a narrower band than a single strip. This graph shows that the off-diagonal components of the **Z** matrix are critical to creating the full CPL structure and act as a way of filtering out certain frequencies. Since the receiving antenna has the same geometry as the input antenna and is governed by the same **Z** matrix, it will have the same frequency band during reception and will perform the same filtering a second time, thus further narrowing down the transmission band.

### 3.2.4. Spin wave signal formation and reception of $S_{21}$ and $S_{12}$

The detection of spin waves by the receiving antenna is a consequence of the electric field of the propagating spin waves (Eq. (41)) at the location of the output antenna. Let us first look at the electric field of a single infinitely thin wire, given by Eq. (22), as a function of the distance $x$ from the wire, shown in Figure 10. This is calculated for a frequency of 17.886 GHz, which corresponds to the maximum of spin-wave transmission in the $+x$ direction for the default system parameters.

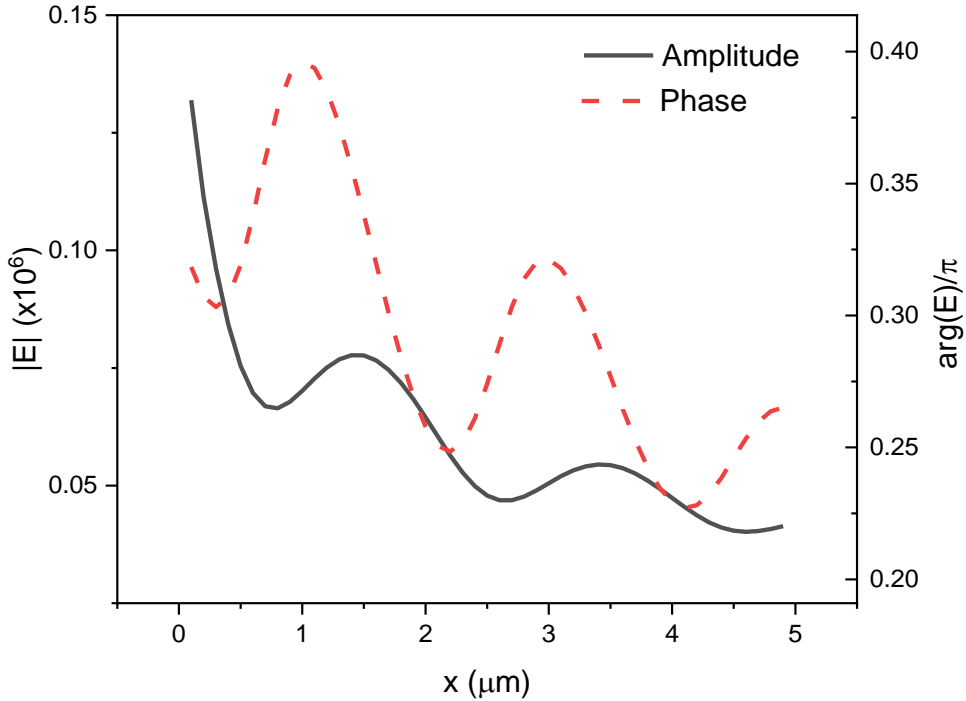

*Figure 10: Black-solid line: Amplitude of the electric field of a spin wave excited by an infinitely thin wire carrying a microwave current. Red-dashed line: Phase of the electric field.*

The amplitude of the electric field strength obeys an exponential decay which is governed by the magnetic losses in the ferromagnetic layer. Additionally one notices an oscillatory background which is attributed to the interference of the electric field of the spin wave with the electric field induced in the vicinity of the antenna by the Oersted field of the microwave current in it [69]. The latter signal is usually responsible for the direct inductive coupling of the input and output antennas and is described by Eq. (S32) of Ref. [25]. Therefore, in the following, we will term it "the direct inductive coupling". The phase of the electric field resembles a linear behavior with the addition of a large oscillatory background. The complex behavior of the phase is similarly a result of the interference of the spin-wave signal and the direct inductive coupling. This can be described as follows: let us approximate the spin wave as an exponentially decaying wave in the spatial direction $x$,

$$E_{sw} = Ae^{-\alpha x}e^{-ikx}, \qquad (48)$$

where $A$ is a real-valued amplitude constant, $\alpha$ is the exponential decay constant of the spin wave, and $k$ is the wavenumber of the propagating spin wave. Now, assuming the input antenna is an infinitely thin wire with infinite length in the $z$-direction, carrying a current $I$, then we may approximate the contribution of the direct inductive coupling as,

$$E_{ind} \approx \frac{I}{2\pi x} \sim \frac{h}{x}. \qquad (49)$$

(Here we neglected the conductivity of the layered structure that is supposed to screen $h$.) We expect that this signal is amplified by magnetization precession, more precisely by its part that does not propagate as a spin wave, but just follows the local microwave $h$ field. This is the near field of the antenna [41]. To account for this effect, we multiply $E_{ind}$ by the microwave magnetic susceptibility $\mu$. Ultimately the direct-inductive-coupling contribution becomes,

$$E_{nf} \approx \frac{\mu I}{2\pi x} \sim \frac{C}{x}, \qquad (50)$$

where $C$ is a real-valued amplitude constant and $\mu$ is the scalar microwave magnetic susceptibility of the magnetic material [57].

The total electric field expected at some distance $x$ from the input antenna thus becomes,

$$E_{tot} = Ae^{-\alpha x}e^{-ikx} + \frac{C}{x}. \qquad (51)$$

Calculating the amplitude of the electric field is then straightforward and one obtains,

$$|E_{tot}| = Ae^{-\alpha x}\sqrt{1 + 2\cos(kx) f(x) + f(x)^2}, \qquad (52)$$

where $f(x) = \frac{C}{A}\frac{e^{\alpha x}}{x}$. The amplitude obeys an exponential decay overlaid by an oscillatory function whose period is determined by the wavenumber $k$.

The phase of the electric field is obtained via,

$$\arg(E_{tot}) = \mathrm{atan}\left(\frac{-Ae^{-\alpha x}\sin(kx)}{Ae^{-\alpha x}\cos(kx) + \frac{C}{x}}\right). \qquad (53)$$

When both coupling terms are negligibly small such that $C \ll 1$ then the amplitude of the electric field is simply given by the amplitude of the spin wave, $|E_{tot}| = Ae^{-\alpha x}$. In addition, the phase of the electric field, in this case, reduces to the linear relation, $\arg(E_{tot}) = -kx$, which is the phase accumulation of the propagating spin wave. When the coupling terms are non-negligible, we return to the non-trivial case where the electric field amplitude has an additional oscillating term and the phase no longer follows a linear relation.

Let us now consider the full electric field of the spin waves excited by the CPL (Eq. (41)). Figure 11 shows the amplitude and phase of the electric field as a function of the distance in the positive propagation direction, $+x$, at a microwave input frequency of, 17.886 GHz. This electric field corresponds to a spin wave with a positive wavenumber and excited by the coplanar antenna. Again one notices oscillations in the amplitude of the electric field. These stem from the same interference of the spin-wave signal with the near field of the input antenna. From the figure, one sees that the oscillations of the amplitude are now smaller, and the phase is almost linear. This demonstrates that the near field of a CPL is much more strongly localized near the excitation antenna than the near field of a single wire, such that for distances $4\mu$m or above the signal carried by the traveling spin wave dominates.

This is one important advantage of using coplanar antennas – the joint action of the input and output CPL antennas efficiently filters out the parasitic direct-inductive-coupling signal that would be much stronger otherwise (for instance, if microstrip antennas were used instead). The output signal then originates mostly from the electric field of the spin wave. The contribution of the output antenna to the process is as follows. A useful signal carried by a CPL is a signal for which currents in the CPL's signal and ground lines are in anti-phase. Only this type of signal can be picked up by a device or instrument sitting at the end of a CPL, e.g., a microwave vector network analyzer (VNA). However, the electric field stemming from the direct inductive coupling has a constant phase in the spatial dimension and the resultant induced currents from this electric field remain in phase across all three strips of the output antenna. Thus, these currents do not contribute to the useful signal at the output port. On the other hand, the electric field of the propagating spin wave does accumulate phase in space and since the input and output antennas have the same spatial geometry, the excited spin wave from the input antenna will couple efficiently to the detection band of the output antenna and induce anti-phase currents between the signal and ground strips at the output antenna. These currents will contribute to the useful signal at the output port.

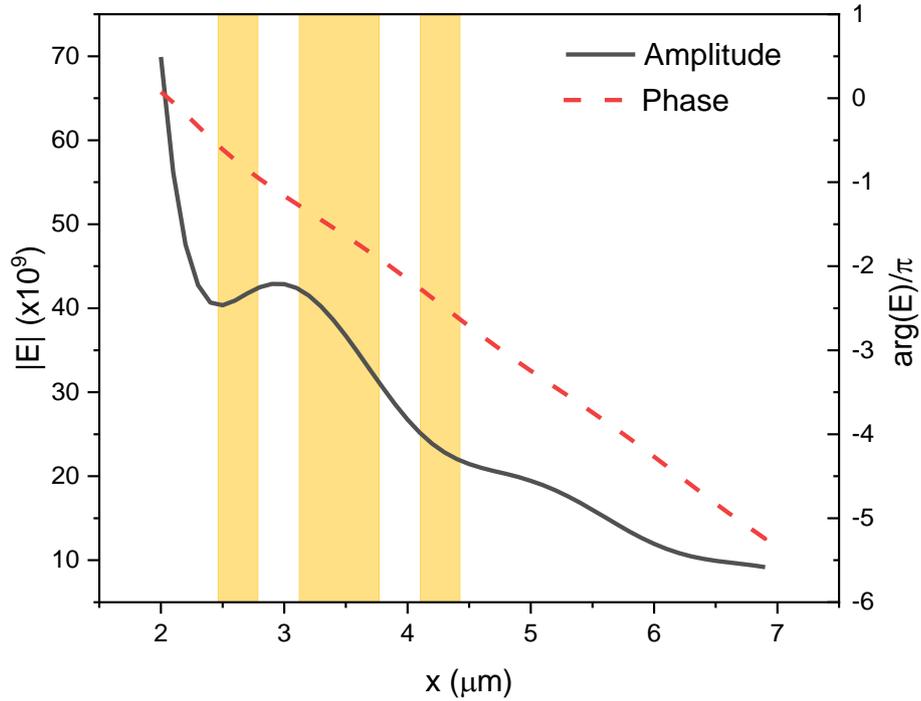

*Figure 11: Black-solid line: Amplitude of the electric field of the propagating spin wave excited by a CPL antenna. Red-dashed line: Phase of the electric field of the spin wave. The electric field was calculated for an incident input frequency of 17.886 GHz, which corresponds to the maximum spin-wave transmission in the +x direction. The gold rectangles indicate the locations of the strips of the receiving CPL antenna.*

### 3.2.5. iPMA, iDMI, $4\pi M_s$, and Gyromagnetic Ratio

Lastly, we investigate how the various magnetic parameters of the system affect the transmission characteristic as a result of the changes in the dispersion relation of the spin waves. The reader will find this discussion in Section 7 of the online supplemental materials [25] to this article.

## 4. Conclusion

A fully self-consistent model for the excitation and reception of magnetostatic surface waves (MSSW) by a set of coplanar (CPL) antennas was developed and implemented numerically. The model assumes that the ferromagnetic film is conducting and is interfaced with non-magnetic metallic films, but is also suitable for modeling magneto-insulating films. Interface perpendicular magnetic anisotropies and Dzyaloshinskii-Moriya interaction are introduced at both interfaces of the ferromagnetic layer.

In addition, our model includes the contribution of inhomogeneous exchange interaction to SW energy, properly accounts for intrinsic Ohmic resistance of the antenna strips by including the resistivity directly into the derived integral equations, and allows the ground lines of the coplanar antennas to have arbitrary finite widths. We cast the solution of the obtained integral equations into an inductive-impedance tensor. A system of five Telegrapher Equations was then derived that employs the tensor. In this way, the Telegrapher Equations accounted for the effect of non-reciprocity of spin-wave excitation and reception by the coplanar antennas.

The final output of the model is the scattering $S$-parameters, $S_{11}$, $S_{22}$, $S_{12}$, and $S_{21}$ of the CPL antenna system. These were calculated numerically and their respective amplitude and phase profiles as a function of microwave frequency were in good qualitative agreement with what has been observed in previous experimental studies [14,17,19,22,23].

An interesting finding is that in the absence of any asymmetry of the spin-wave dispersion relation there remains a small frequency non-reciprocity between the forward and backward propagating spin waves as calculated by the $S_{21}$ and $S_{12}$ parameters respectively. A detailed analysis of theoretical tests that led to this finding was outlined in section 3.2.2. Such a frequency non-reciprocity has not been found in previous theoretical studies employing the self-consistent model [30,32,36,38], and thus stems from our more in-depth formalism of the impedance wherein we calculate the full cross and mutual couplings of the individual antenna strips utilizing a 3x3 impedance matrix.

In addition, the intermediate results of the model leading to the final $S$-parameters were checked numerically and discussed in the study. These intermediate results correspond to the electric field of a MSSW excited by an infinitely thin wire in the Fourier and real space, the electric field of a MSSW excited by a CPL antenna in real space, the current distribution in the excitation antenna and its corresponding Fourier transform, and the input impedance of a single strip antenna and CPL antenna in the frequency space.

Lastly, the flexibility of the model allows one to introduce surface perpendicular magnetic anisotropy (PMA) and interfacial Dzyaloshinskii-Moriya interaction (iDMI) at both interfaces of the ferromagnetic metal.

Furthermore, numerical calculations showed that our developed model is in good agreement with well-established models for the spin-wave dispersion relation of MSSWs. Finally, the model developed in this study was used to provide fits for experimental propagating-spin-wave spectroscopy measurements carried out on a series of //Ru(5)/Co(x)/Pt(5) (x = 2, 5, 10, 20 nm and numbers in parenthesis are given in nm) samples, as discussed in the second paper of this series of papers [70].

Finally, it may be worth mentioning that our model may also be extended to modeling meander CPL antennas. To model an antenna comprising $n$ half-periods of a meander, one will need to obtain a **Z** matrix for $3n$ parallel strips and derive a new system of Telegrapher Equations that will now consist of $5n$ equations. This is necessary because the half-periods of the meander will be coupled both through the usual mutual inductance but also by partial spin waves excited by each half-period. The last step will be to "stitch" the half-periods by applying appropriate boundary conditions for currents and voltages and solve the system of Telegrapher Equations.

Acknowledgments

A Research Collaboration Award from the University of Western Australia (UWA) is acknowledged. C.W. acknowledges his Research Training Program stipend from UWA and a travel grant from the Franco-Australian Hubert Curien Program.

## 5. Justification for utilizing MSSW's

Consider the group velocity of spin waves $V_g$. From the approximate expressions in [1], in the limit of small wavenumbers and $H, H_i \ll M_0$ we obtain that the group velocity scales as $LH_i$ for Forward Volume Magnetostatic Spin Waves (FVMSW), $L\sqrt{MH}$ for Backward Volume Magnetostatic Spin Waves (BVMSW) [2] and $L\sqrt{M^3/H}$ for MSSW. (Here $H$ is the applied field, $H_i$ is the internal field of a film, $M_0$ is the film's saturation magnetization, and $L$ is the film thickness). Given the linear scaling with $L$, we expect very small group velocities of SW for FVMSW and BVMSW in samples with thicknesses in the nanometer range and reasonably small values of the applied/internal field. Conversely, because of the scaling as $1/\sqrt{H}$, MSSW group velocity appears to be large for $H \ll M_0$, and especially for materials with large saturation magnetization values such as cobalt, iron, nickel, and their alloys because of the scaling as $\sqrt{M^3}$.

The group velocity is important because it determines how far a wave may travel once excited with a stripline antenna before it dies off due to magnetic losses in a medium. Put differently, the length of the free-propagation path of spin waves scales as $V_g$. A wave may be of practical interest only if it can leave the area of its excitation before it dies off. This implies that the free propagation path must be significantly larger than the antenna size in the direction of propagation of the spin wave excited by the antenna. (The size is usually called "antenna width".) This suggests that only the MSSW is of practical interest for applications unless we are able and it is reasonable to apply very large magnetic fields to a film. Furthermore, the excitation of MSSWs by stripline antennas is unidirectional [3], which is also very useful for applications. For this exact reason, experiments from the literature exploited MSSW [3–14].

## 6. Exchange boundary conditions and their effect on magnetization dynamics

Note that the exchange boundary conditions may affect the magnetization dynamics only if the film thickness is small enough such that the contribution of the film thickness to the effective field of the inhomogeneous exchange interaction is comparable to or larger than the dynamic dipole field, which the precessing magnetization vector induces. Then, from Eq. (2) it follows that it must hold that $\alpha/L^2$ is on the order of 1 or larger in order for the exchange interaction to affect the dynamics and make $q_i$ noticeably different from $\pm|k|$, which is characteristic to the non-exchange (more precisely, dipole-dominated) MSSW. This is always the case for metallic ferromagnetic layers with sub-skin-depth thicknesses (below 150 nm or so). In this case, surface/interface anisotropies and interface exchange, including iDMI, at surfaces/interfaces of the ferromagnetic layer may affect magnetization dynamics of the bulk of the layer, because of the strong contribution of the exchange interaction to the collective dynamics of spins. This does not happen if the dipole coupling dominates because it is a long-range interaction and, as such, can smear out any significant variations in the precession amplitude if the length scale of the precession amplitude non-uniformity due to local magnetic inhomogeneities, e.g. interface anisotropy, is significantly smaller than the length scale for dipole coupling.

Figure S1 illustrates the effect of the exchange boundary condition on the magnetization dynamics. It shows two profiles of MSSW in a Co layer with a thickness of 20 nm for $K_u = 0$ and $K_u = 4.24 \times 10^{-3}$ $J/m^2$. The profiles were calculated for $k=0$ with the numerical model from [15]. One sees that the presence of the interface PMA does not affect the in-plane component $m_x$ of dynamic magnetization, but significantly increases the amplitude of the perpendicular-to-plane component $m_y$ near the layer interface where it is present. The overall significantly smaller amplitude of $m_y$ hints at the reason for the increase in the amplitude at the interface. The large difference in the overall magnitudes of the two components is due to a strong dynamic demagnetizing field in the y-direction. For $k=0$, this field equals $-m_y$, and there is no in-plane dynamic demagnetizing field component. The presence of the large perpendicular-to-plane demagnetizing field squashes the magnetization-vector precession cone in the y-direction – $m_y$ becomes significantly smaller than $m_x$. The interface PMA, as a uniaxial anisotropy, acts in the opposite direction – it tries to stretch the cone along the y-axis. Due to the interface character of PMA, this only happens at the interface. The profile "deformation" propagates into the bulk of the film due to the exchange coupling of bulk spins to the interface ones.

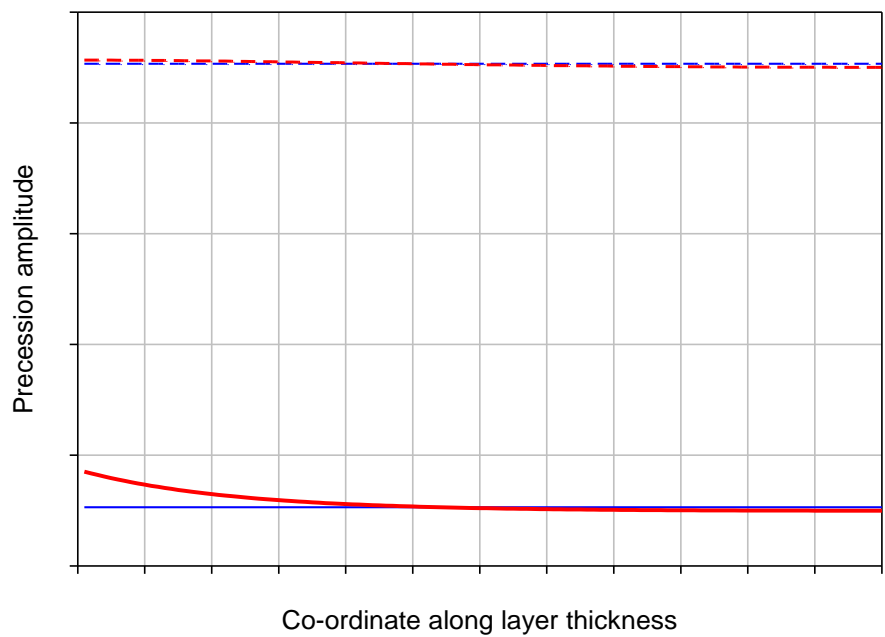

*Figure S1: Amplitude of the $m_y$ and $m_x$ components of dynamic magnetization (solid and dashed lines respectively) as a function of co-ordinate along the thickness of the ferromagnetic layer. Thick red lines: vanishing interface PMA at the film surface y=0 (left edge of the graph). Thin blue lines: interface $K_u = 4.24 \times 10^{-3}$ $J/m^2$.*

## 7. Solving the Telegrapher Equations for appropriate boundary conditions

The vectors in Eq. (31) of the main text are given by,

$$|b> = \begin{vmatrix} I_1 \\ I_2 \\ I_3 \\ V_{12} \\ V_{13} \end{vmatrix}, \quad (S1)$$

$$|F> = \begin{vmatrix} 0 \\ 0 \\ 0 \\ \Delta U_2 - \Delta U_1 \\ \Delta U_3 - \Delta U_1 \end{vmatrix}, \quad (S2)$$

and $\hat{A}$ is the matrix of coefficients of the system

$$\hat{A} = \begin{pmatrix} 0 & 0 & 0 & -Y_{12} & -Y_{12} \\ 0 & 0 & 0 & Y_{12} & 0 \\ 0 & 0 & 0 & 0 & Y_{12} \\ (Z_{11} - Z_{21}) & (-Z_{12} + Z_{22}) & (Z_{23} - Z_{13}) & 0 & 0 \\ (Z_{11} - Z_{31}) & -(Z_{12} - Z_{32}) & (-Z_{13} + Z_{33}) & 0 & 0 \end{pmatrix}. \quad (S3)$$

Equation (31) admits a general solution as follows

$$|b(z)> = \hat{X} | diag\left(\exp(\gamma_n z)\right) | b^0 > + \hat{X} \, diag\left(\frac{\exp(\gamma_n z) - 1}{\gamma_n}\right) \hat{X}^{-1} | F >. \quad (S4)$$

Here $\gamma_n$ is the *n*-th eigenvalue of $\hat{A}$, (*n*=1,2,..5), $\hat{X}$ is a matrix composed of eigenvectors of $\hat{A}$ in such a way that its *n*-th column represents the eigenvector that corresponds to the *n*-th eigenvalue, $diag(s_i)$ is a diagonal matrix that has the expression $s_i$ as its (*i,i*) element, and |$b^0$> is an arbitrary vector.

The five eigenvectors of $\hat{A}$ represent five eigenmodes of the coplanar-line structure. There are two pairs of propagating modes (±|$\gamma_n$|≠0), and a mode of direct current ($\gamma_n$=0). The propagating modes are a coplanar-line-like mode with the microwave currents in the signal line and ground lines being in anti-phase and a slot-line-like mode with zero current in the signal line and the currents through the ground lines being in anti-phase to each other. We may expect that the contribution of the slot-line-like mode to the total solution not to vanish on the frequency band of spin-wave excitation by the antenna, because of the strong non-reciprocity of excitation of MSSW, but to vanish elsewhere.

In order to determine $|b_0>$, we need to apply boundary conditions at $z=0$ (the far end of the antenna) and $z=l_s$ (the antenna ports), see Figure 4 in the main text. For the input antenna, $|F>=0$, and Eq. (S4) reduces to

$$|b(z)> = \hat{X} | diag(\exp(\gamma_n z))|b^0>. \tag{S5}$$

We assume that the far end of the input antenna ($z=0$) is a short end. This implies that

$$V_{12}(z=0) = 0; V_{13}(z=0) = 0, \tag{S6}$$

and from the Kirchhof law for currents at a node, we have

$$I_1(z=0) + I_2(z=0) + I_3(z=0) = 0. \tag{S7}$$

We also assume that the input port of the antenna $z=l_s$ is connected to a regular coplanar line with a characteristic impedance $z_c$ of 50 Ohm that is fed by a source microwave voltage $V_0$. The respective equivalent circuit is shown in Figure 4. From this circuit, it follows that

$$V_{12}(z=l_s) + 2z_c I_2(z=l_s) = 2V_0, \tag{S8}$$

and,

$$V_{13}(z=l_s) + 2z_c I_3(z=l_s) = 2V_0. \tag{S9}$$

Note the factors of 2 in the expression. The one in front of $z_c$ is there because the partial characteristic impedance between the signal line and any of the ground planes of a coplanar line is $2z_c$. The two partial impedances (between the signal line and the ground line 1 and between the signal line and the ground line 2) are connected in parallel; therefore, the total characteristic impedance of a coplanar line is $z_c$. The one in front of $V_0$ is because we are dealing with a microwave circuit (see e.g. [16]).

Eqs. (S8)-(S9) are the last two boundary conditions needed to determine the five components of $|b_0>$. The first three components of the vector $|b(z)>$ (Eq. (S5)) are currents in the signal and ground lines that enter Eqs. (S7)-(S9), and the last two components are the voltages from Eqs. (S8)-(S9). Substitution of Eq. (S5) into Eqs. (S6)-(S9) yields a system of equations

$$\sum_{i=1}^{3}\sum_{j=1}^{5} X_{i,j} b_j^0 = 0$$

$$\sum_{j=1}^{5} X_{4,j} b_j^0 = 0$$

$$\sum_{j=1}^{5} X_{5,j} b_j^0 = 0 \qquad , \qquad (S10)$$

$$\sum_{j=1}^{5} (X_{4,j} + 2z_c X_{2,j}) \exp(\gamma_j l_s) b_j^0 = 2V_0$$

$$\sum_{j=1}^{5} (X_{5,j} + 2z_c X_{2,j}) \exp(\gamma_j l_s) b_j^0 = 2V_0$$

where $X_{i,j}$, is the $i$-th component of the eigenvector of the matrix $\hat{A}$ (Eq. (S3)) that corresponds to its $j$-th eigenvalue, and $b_j^0$ is the j-th component of the vector $|b_0\rangle$. This system of equations allows a simple numerical solution using numerical methods of Linear Algebra. Once the system has been solved, we can calculate partial input impedances of the current loops shown in Figure 4 of the paper. The partial impedances are given by Eqs. (32) and (33) in the main paper.

## 8. Justification for neglecting z-dependence of current in antennas

The input currents of the input antenna are given by the first three components of the solution vector (Eq. (S1)) evaluated at $z=l_s$. As seen from Eq. (S4), the currents vary along the strips. Therefore, the amplitude of the excited spin wave is not constant along the spin-wave wavefront (i.e. in the z-direction). This idea is easier to understand in the case of a single-strip (i.e. microstrip antenna). The antenna is short-ended, therefore the microwave current in it scales as $\exp(\gamma z) + \exp[\gamma(l_s-z)] = \cosh[\gamma(z-l_s/2)]\exp(\gamma l_s/2)$, where $z=l_s$ is the co-ordinate of the short-end, and $\gamma$ is the propagation constant that is a single-strip equivalent of $\gamma_n$ from Eq. (S4). Note that $\gamma$ is complex-valued. This expression physically means that we have a standing wave of current with its anti-node at $x=l_s$. Consider two limiting cases. The first one is for no Ohmic losses in the antenna, with the antenna being very weakly loaded by the spin wave excitation. Then $\gamma$ is purely imaginary ($\gamma=i\cdot\mathrm{im}(\gamma)$), and we have the current amplitude decreasing parabolically from the short end. The phase of the current is the same everywhere, as the original sum of the exponents reduces to $\cos[\mathrm{im}(\gamma)(z-l_s/2)]$. This expression describes a standing wave of microwave current in the antenna of current with an anti-node at $z=l_s$. Conversely, if Ohmic losses in the antenna dominate we have $\gamma=\mathrm{Re}(\gamma)<0$ and $\cosh[\gamma(z-l_s/2)]\exp(\gamma l_s/2)$. For very large losses $\gamma l_s \ll -1$, this expression reduces to $\exp(-|\gamma|z)$. Hence, we have a current amplitude decaying exponentially from the antenna input to its short end. Again, the current amplitude is not uniform along $z$. Then in the general case of an essentially complex-valued $\gamma$ we may expect both amplitude and phase of the current to vary along z.

Each point $z$ of an antenna is a point source of spin waves. In the case of MSSW, these are caustic waves [17]. These waves interfere to form the wavefront of a wave excited by the antenna. (The wavefront is a line, along which the wave phase is constant.) If the amplitude and phase of the current were constant along the antenna, just from the symmetry reason, we would expect a wavefront of the excited wave to be perfectly parallel to the antenna and the magnetization precession amplitude to be the same all along the front. As such, the wave would propagate perfectly perpendicular to the antenna and reach the output antenna with its front perfectly perpendicular to it. Then, ignoring losses to diffraction during the propagation, one would have the whole power of the incident wave reaching the output antenna and the same phase of the incident wave all along the output antenna. We may neglect the diffraction if the distance between the antennas is smaller than the antenna length $l_s$. This condition is usually satisfied for microscopic antennas [3–14].

However, if the current is not uniform along the antenna, we cannot expect that the interference process produces a perfectly linear wavefront. The wavefront will be more complicated, with caustic behavior present in it. Accordingly, the process of wave propagation to the output antenna is essentially two-dimensional in the general case and cannot be reduced to a 1D problem. However, earlier theories of macroscopic spin waves in YIG films with micrometer-range thicknesses implicitly assumed that MSSW wavefronts are perfectly parallel to antennas and that the precession amplitude is perfectly uniform along the wavefront [12,18–30]. These theories are in good agreement with experiments, except for a correction coefficient of $2/\pi$ for the antenna excitation efficiency found experimentally in [31]. The good agreement is due to the fact that all experimental studies employed films of finite sizes along the antenna. This resulted in a standing-wave pattern along that direction and propagation of the wave as a guided wave. The wavefront of the lowest-order width mode [32] of guided MSSW is parallel to the strip waveguide by definition. The correction coefficient then can be explained as due to some non-uniformity of magnetization precession along the wavefront because of the formation of a standing wave in this direction.

The situation is different for MSSW in metallic magnetic films with nanometer-range thicknesses. The typical length of the antenna and width of the film strip in the antenna direction is comparable to or larger than the free-propagation path for the MSSW. Therefore, we do not expect a standing-wave MSSW pattern to form along $z$. Rather, we must consider each point of the antenna to be an independent source of caustic waves and solve the problem of MSSW beam formation two-dimensionally. To avoid this difficulty, we assume that the antenna is short with respect to the wavelength of the wave of microwave current in it $|\gamma|l_s<<1$. In this limit, we may expand the current dependence on z into Taylor series $\exp(\gamma z) + \exp[\gamma(l_s-z)]\approx 2+\gamma l_s+ \gamma^2(z^2- l_sz)$. Ignoring the first- and the second-order terms, we obtain that the current is uniform along the antenna.

The approximation $|\gamma|l_s<<1$ must work fine in the frequency range below ~30 GHz. For this range, the wavelength of an electromagnetic wave in free space is larger than 1 cm. Even if the wavelength is reduced due to the presence of a dielectric substrate of a stripline antenna we do not expect the wavelength to become shorter than 1 mm. This is still much larger than the antenna lengths from the known experiments [3–14] – on the order of tens of microns. The same observation remains valid if we also take into consideration reasonable Ohmic losses in the antenna.

Then, combined with the fact that typical distances $l_d$ between the input and output antennas are smaller than $l_s$ [4], the same 1D approximation of MSSW wavefronts parallel to the antennas and precession amplitude perfectly uniform along the wavefront and for thick YIG films must be valid for the microscopic antenna geometries.

## 9. Details of the derivation of the output voltage of the output antenna

The characteristic impedance for the CPL section is $z_c$ (e.g. 50 Ohm). This implies that for the quantities in Figure 3, we now have

$$V_{12}(z = l_s)/I_2(z = l_s) = 2z_c, \qquad \text{(S11)}$$

$$V_{13}(z = l_s)/I_3(z = l_s) = 2z_c, \qquad \text{(S12)}$$

where $I_i(z)$ is now given by $I_i^{\text{out}}(z)$, and $V_{ij}(z)$ denotes voltages in the output antenna. Note the factor of 2 in front of $z_c$. We explained this factor while discussing Eq. (S9).

The boundary conditions at $z=0$ are the same as for the input CPL. Application of the boundary conditions to Eq. (S4) allows us to easily find the vector $|b_0\rangle$ for the output antenna. This is straightforward linear algebra under the assumption that $\Delta U_i$ is uniform along $z$ because the antenna is short (see the discussion above Eq. (48)); therefore, we omit the calculation here. Once $|b_0\rangle$ has been obtained, one easily finds $V_{12}(z = l_s)$ and $V_{13}(z = l_s)$. These are the output voltages for the section of a regular CPL.

The output voltages are given by the expression

$$V_{1i} = \sum_{j=1}^{5} X_{i+3,j} c_j \exp(\gamma_j l_s) + B_{i+3}, \qquad \text{(S13)}$$

where $i = 2$ or $3$, $X_{ij}$ are components of the matrix $\hat{X}$, and the column vector $B$ is given by the second term of Eq. (S4):

$$B = \hat{X} \, diag\left(\frac{\exp(\gamma_n z) - 1}{\gamma_n}\right) \hat{X}^{-1} | F \rangle. \qquad \text{(S14)}$$

The quantities $c_j$ are components of a column vector $\boldsymbol{c} = \hat{C}^{-1} \boldsymbol{p}$, where the column vector $\boldsymbol{p} = (0, 0, 0, p_4, p_5)$ with components $p_i = B_i + 2\, B_{i-3}\, z_c$, ($i = 4, 5$). The matrix $\hat{C}$ has components as follows:

$$C_{1i} = X_{1i} + X_{2i} + X_{3i}, C_{2i} = X_{4i}, C_{3i} = X_{5i}, C_{4i} = (2X_{2i} z_c + X_{4i}) \exp(\gamma_i l_s). \quad \text{(S15)}$$

and

$$C_{5i} = (2X_{3i}z_c + X_{5i})\exp(\gamma_i l_s), \text{ where } i = 1, 2, \ldots 5. \tag{S16}$$

The CPL section is usually connected to a standard feeding line. For instance, this may be a coaxial cable that feeds the output voltage into a vector network analyzer. Alternatively, this may be a microscopic microwave probe ("Picoprobe") connected to the contact pads of the CPL section. Importantly, in the process of making the connection, both ground lines of the microscopic CPL become connected to the same ground line of the feeding line (e.g. to the shield of the coaxial cable). The signal line of the CPL becomes connected to the signal line of the feeding line (e.g. to the core of the coaxial cable). As a result, the two voltage sources – $V_{12}$ and $V_{13}$ – become connected in parallel. This yields a voltage $V_{out}$ at the input of the feeding line (e.g. between the core and the shield of the coaxial cable)

$$V_{out} = (V_{12} + V_{13})/2, \tag{S17}$$

## 10. Comparing theoretical simulation with the analytical dispersion relation

Let us now convert the frequencies of the maximum of transmission and minimum of reflection into their corresponding wavenumbers via the dispersion relation of the system. The dispersion relation for the MSSWs follows [17],

$$\omega(k) = \sqrt{[\omega_H + \omega_M(\alpha k^2 + 1 - N_u - P)][\omega_H + \omega_M(\alpha k^2 + P)]} + \frac{2\gamma k D}{LM_s}, , \tag{S18}$$

where $\omega$ is the frequency of the spin wave, $\omega_H = \gamma\mu_0 H$, $\omega_M = \gamma\mu_0 M_s$, H is the applied field, $M_s$ is the saturation magnetization of the ferromagnetic layer, $\gamma$ is the gyromagnetic ratio, $\alpha = 2A/(\mu_0 M_s^2)$ is the exchange constant, D is the total surface DMI constant, $L$ is the thickness of the ferromagnetic layer, and $k$ is the wavenumber of the MSSW. Lastly, $N_u = 2K_u/(M_s^2 L)$ and $P = 1 - [1 - e^{-|k|L}]/(|k|L)$, where $K_u$ is the total surface PMA constant. Note that both $D$ and $K_u$ may have contributions from both interfaces. Note that the analytical Eq. (S18) is approximate. It is known to give a good approximation for the dispersion relation. However, if one wishes to more accurately model the surface PMA and DMI contributions it is beneficial to use a more accurate theory of dispersion (see e.g. Ref. [4]).

To this end, we developed a numerical model which breaks the thickness of the ferromagnetic layer into discrete mesh steps and solves the Landau-Lifschitz equation at each mesh point. Both dipole-dipole and exchange contributions to the dynamic effective magnetic field were accounted for. The surface PMA and surface DMI contributions were included as bulk effective fields, but acted only at the mesh points closest to each interface. The conductivities of the metal layers are not included in the model. It is well known experimentally that BLS spectra for metallic magnetic films are in good agreement with spin-wave dispersion relations that do not include the conductivities of magnetic films. This is because the conductivity mostly affects spin waves with wavelengths larger than the skin depth for the film material, provided the film thickness is much smaller than the skin depth. This wavelength range is a very small part of the wavelength range that BLS can access. Furthermore, the same applies to the limiting case of the ferromagnetic resonance (a spin wave with $k$=0), for which a non-negligible film conductivity affects the strength of the FMR response of the film, but not the FMR peak position [33]. The same is valid for propagating spin-wave spectroscopy employing CPL antennas – we expect that the finite conductivity of the metallic film influences the efficiency of excitation and reception of spin waves by the antennas, but not the dispersion law for the excited spin waves. Using this numerical approach allows for careful tailoring of the individual surface constants at each interface and does not average these surface contributions over the entire bulk of the sample as is the case in Eq. (S18). The dispersion relation obtained for the system with the default parameters using the numerical model is shown in Figure S2. To obtain the dispersion for the negative $k$ values we simply change the sign of the magnetic field and the direction of the saturation magnetization.

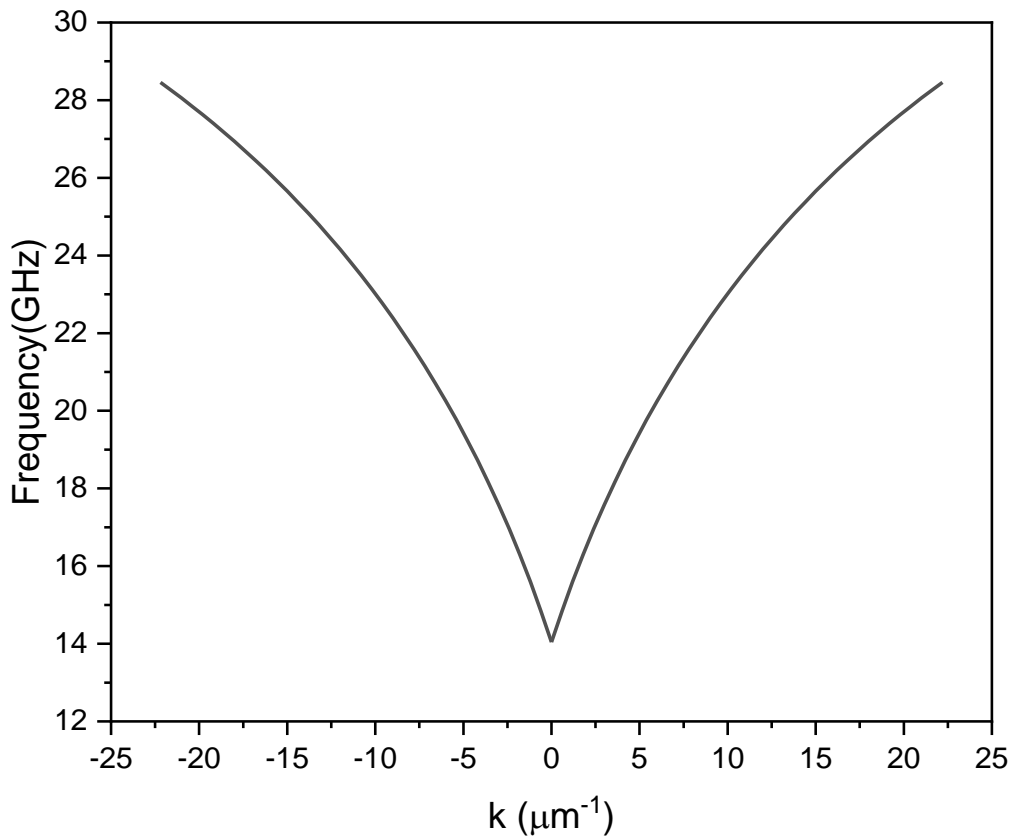

*Figure S2: Dispersion relation obtained from the numerical model using the default system parameters.*

Using the resulting dispersion relation in Figure S2, the wavenumbers of maximum transmission for $S_{21}$ and $S_{12}$ are, ~+3.31 $\mu m^{-1}$ and ~-3.41 $\mu m^{-1}$, respectively, and the wavenumber of minimum reflection for $S_{11}$ is ~3.48 $\mu m^{-1}$. The difference between the absolute values of the wavenumbers for the forward and backward traveling waves is 0.1 $\mu m^{-1}$.

## 11. Influence iPMA, iDMI, $4\pi M_s$, and Gyromagnetic Ratio on the SWDL Transmission

Figure S3 shows how the $S_{12}$ transmission characteristic depends on the different magnetic parameters of the system. Increasing the external applied magnetic field, saturation magnetization, or gyromagnetic ratio shifts the spin-wave band to higher frequencies. This is directly confirmed by Eq. (64) as one sees that the frequency scales with all three of these terms. The Gilbert damping constant is directly related to the magnetic losses in the system and thus increased damping will result in a quicker reduction of the amplitude of precession. For a MSSW this is seen as a reduction in the spin-wave amplitude and ultimately results in a reduction of the $S_{12}$ transmission as seen in Figure S3(d).

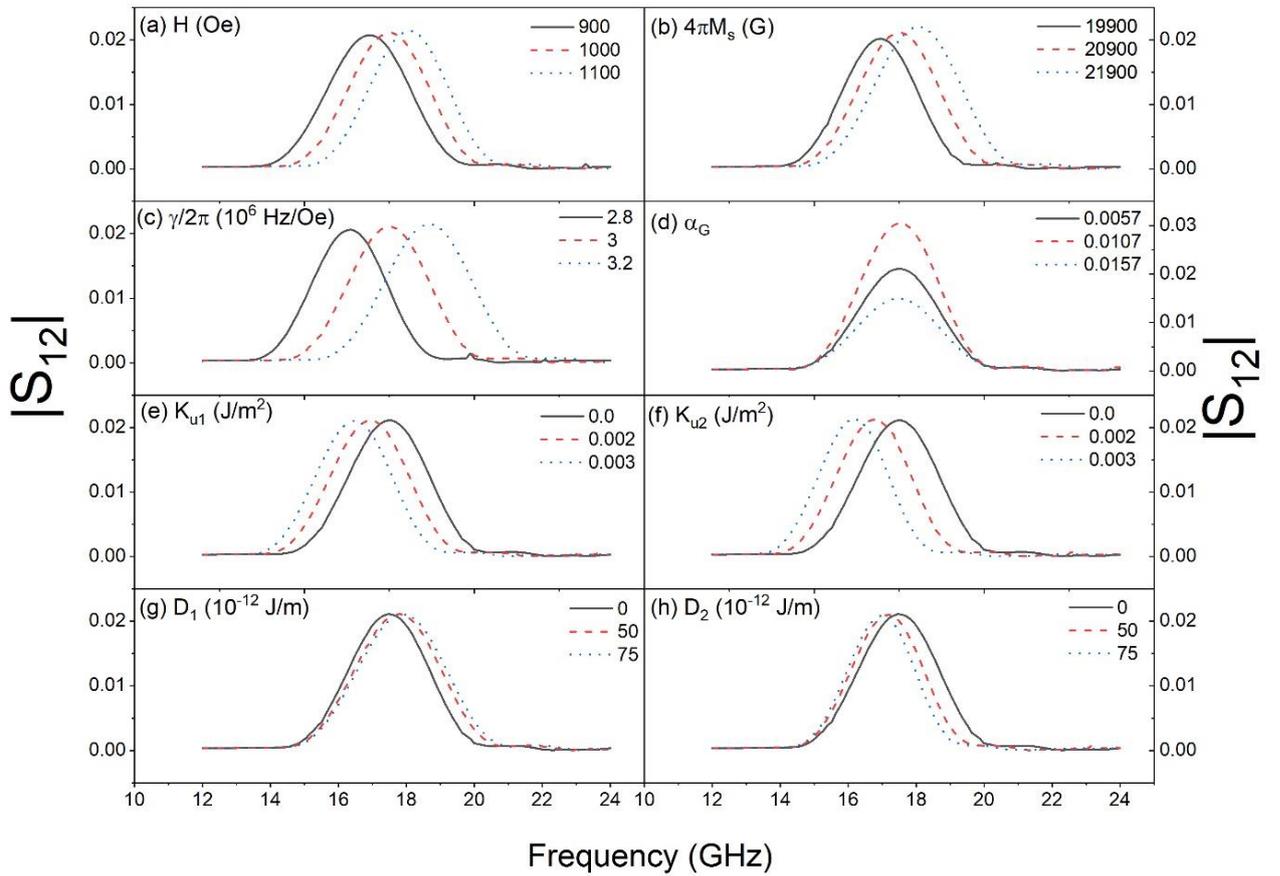

*Figure S3: $S_{12}$ transmission characteristic as a function of input driving frequency for 3 different values of the following system parameters: a) H, b) $4\pi M_s$, c) $\gamma$, d) $\alpha_G$, e) $K_{u1}$, f) $K_{u2}$, g) $D_1$, h) $D_2$. As a reference, the red dashed curve in panels (a)-(d) and the black solid line in panels (e)-(h) is the same and corresponds to the default system parameters.*

In the current formalism, a positive value of the surface PMA constant $K_u$ corresponds to a uniaxial anisotropy. As such, from Eq. (S18) it is evident that a uniaxial PMA contribution will result in a decrease in the spin-wave frequency, which is confirmed by the simulated $S_{12}$ parameter in Figure S3(e,f). For surface PMA the contributions from both interfaces constructively combine to give an overall larger contribution than when considering surface PMA at only one interface. This is because, for uniaxial anisotropy, each interface acts to reduce the spin-wave frequency. Additionally, a frequency non-reciprocity may arise when the surface PMA differs at the two interfaces, which is due to the surface localization of the MSSWs.

Figure S3(g,h) depicts the $S_{12}$ transmission for positive surface DMI constants at the two interfaces respectively. The DMI contribution differs from the PMA contribution in that having the same sign of DMI at the two interfaces results in an overall lower contribution of DMI, as the contributions at the two interfaces counteract each other in this case. This is because the DMI term, the last term in Eq. (S18), depends directly on the spin-wave propagation direction, given by the sign of $k$. In our formalism, opposite signs of DMI at the two interfaces have the same effect on the dispersion, resulting in an overall larger contribution, and the same signs have opposite effects, resulting in an overall lower contribution. Furthermore, since the DMI term depends on the propagation direction, it is clear that the DMI contribution will have the opposite sign for the two MSSW propagation directions resulting in a frequency non-reciprocity.

To confirm the non-reciprocal behaviour of the surface PMA in our model we now investigate the numerical dispersion relation as well as the simulated $S_{21}$ and $S_{12}$ parameters for four cases: 1) $K_{u1} = K_{u2} = 0$ J/m$^2$; 2) $K_{u1} = K_{u2} = 3 \times 10^{-3}$ J/m$^2$; 3) $K_{u1} = 3 \times 10^{-3}$ J/m$^2$, $K_{u2} = 0$ J/m$^2$; 4) $K_{u1} = 0$ J/m$^2$, $K_{u2} = 3 \times 10^{-3}$ J/m$^2$. It is clear from Figure S4(a) that when the surface PMA is the same at both interfaces, there is no frequency non-reciprocity in the dispersion relation which is fully symmetric around the $k = 0$ axis. When the surface PMA contributions differ at the two interfaces then the MSSWs exhibit a frequency non-reciprocity and one obtains an asymmetric dispersion relation around the $k = 0$ axis, as depicted by Figure S4(a). The transmission characteristics $S_{21}$ and $S_{12}$ exhibit similar behavior, albeit with the addition of the non-reciprocity due to spin-wave excitation that is present even in a fully symmetric system as described in the main text. Figure S4(b,c,d,e) show the transmission characteristic, for the four cases, in both propagation directions. The frequency non-reciprocity for each case is shown in the corresponding panel. It is expected that the frequency non-reciprocity is largest for the two asymmetric cases and smallest for the two symmetric cases which is confirmed in Figure S4. When we have no surface PMA at either interface, as shown in Figure S4(b), there should be no frequency non-reciprocity. Hence, the frequency non-reciprocity obtained in this case must stem from a different effect and should be subtracted as a background from all frequency non-reciprocities calculated. This yields the following frequency non-reciprocities, 1) $K_{u1} = K_{u2} = 0$ J/m$^2$, $\Delta f = 0$ MHz; 2) $K_{u1} = K_{u2} = 3 \times 10^{-3}$J/m$^2$, $\Delta f = -12$ MHz; 3) $K_{u1} = 3 \times 10^{-3}$J/m$^2$, $K_{u2} = 0$ J/m$^2$, $\Delta f = -300$ MHz; 4) $K_{u1} = 0$ J/m$^2$, $K_{u2} = 3 \times 10^{-3}$J/m$^2$, $\Delta f = +288$ MHz. As expected, once the background non-reciprocity is removed, the system with symmetric PMA at the two interfaces displays almost no frequency non-reciprocity, whilst the systems with an asymmetric PMA at the interfaces displays a significant non-reciprocity. The sign of the non-reciprocity depends on the interface at which PMA is present due to the forward and backward propagating MSSWs localizing at opposite surfaces of the ferromagnetic layer.

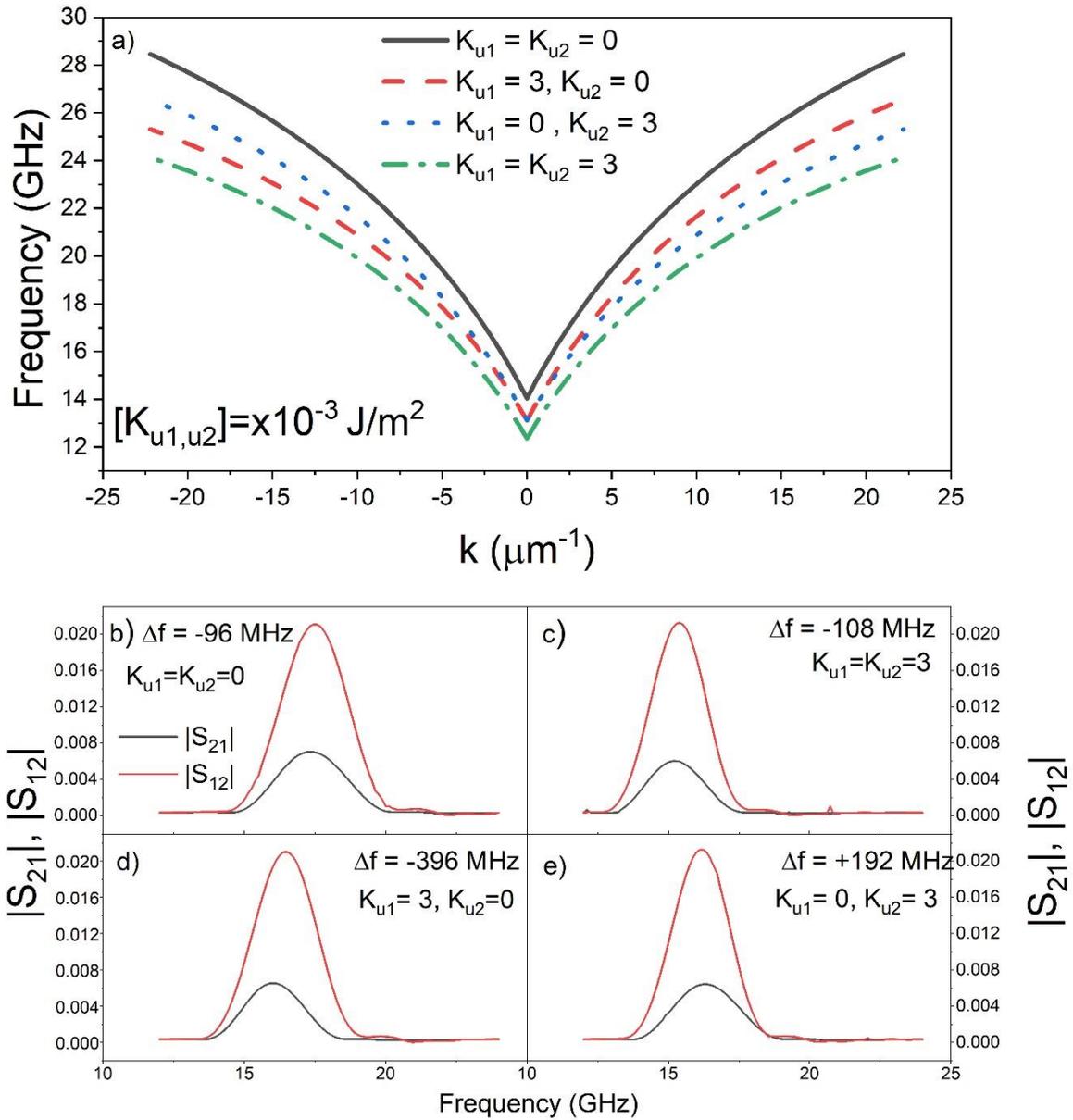

*Figure S4: a) Dispersion relation obtained from the numerical model for the default system parameters and varying anisotropy constants as described in the legend. b-e) Amplitude of the $S_{21}$ (black solid line) and $S_{12}$ (red dashed line) for four different combinations of surface anisotropy constants as shown by the labels for $K_{u1}$ and $K_{u2}$ in the corresponding panels. The surface anisotropy constants are given in $10^{-3}$ J/m$^2$.*

Figure S5 displays how the phase of the $S_{12}$ transmission changes with a change in the magnetic parameters of the system: $H$, $4\pi M_s$, $\gamma$, $\alpha_G$, $K_{u1}$, $K_{u2}$, $D_1$, and $D_2$. The spin-wave band can be seen in the phase profiles as the linear regime. Changing the magnetic parameters may result in a direct translation of the dispersion curve which acts to translate the phase profile in the frequency space, as seen clearly in Figure S5(a, b, c). In addition, changing the magnetic parameters may alter the group velocity of the spin waves resulting in a change in the slope of the dispersion and ultimately a change in the slope of the phase profile, as seen in Figure S5(e, f, g, h).

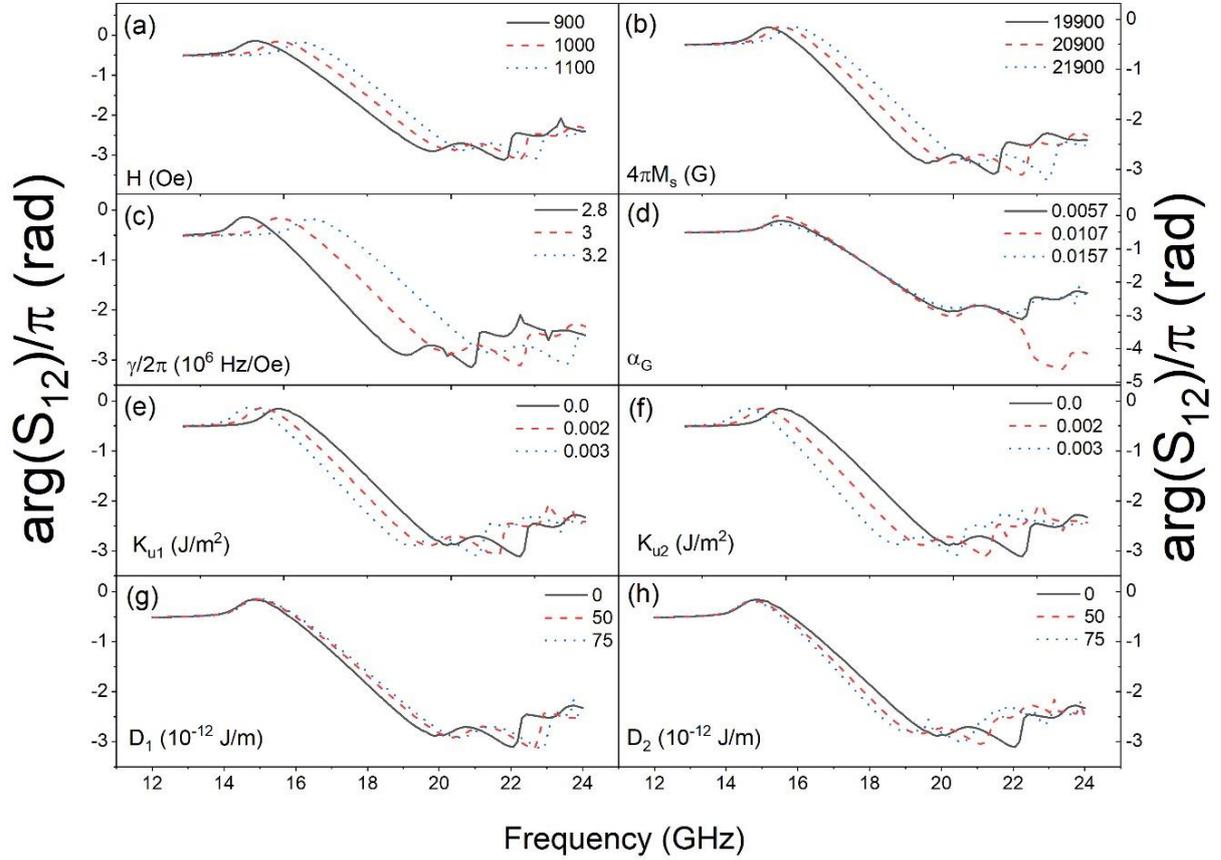

*Figure S5: Phase of the $S_{21}$ transmission characteristic as a function of frequency and how it behaves with various system parameters.*

## 12. Expressions for: $R_m, R_h, C_{my}, C_{hx}, S_0, S_1,$ and $S_j$

The expressions entering Eqs. (18)-(19) of the main text, are as follows,

$$R_m = k^2 \bar{K}^2 / |K|^4, \tag{S19}$$

$$R_h = ik\bar{K}^2 / |K|^4, \tag{S20}$$

where, as before, $K^2 = k^2 + i\mu_0 \sigma \omega$, $\bar{K}$ is the complex-conjugate to $K$, and $\sigma$ is the conductivity of the ferromagnetic layer.

$$S_0 = -i \frac{k}{K_1} \frac{K_1 \cosh(K_1 d_1) + |k| \sinh(K_1 d_1)}{K_1 \sinh(K_1 d_1) + |k| \cosh(K_1 d_1)}, \tag{S21}$$

$$K_1^2 = k^2 + i\mu_0 \sigma_1 \omega, \tag{S22}$$

$\sigma_1$ is the conductivity of the nonmagnetic metallic layer located below the ferromagnetic one ($-d_1<y<0$),

$$S_1 = i\frac{k}{K_2} \frac{K_2 \cosh(K_2 d_2) + |k| \sinh(K_2 d_2)}{K_2 \sinh(K_2 d_2) + |k| \cosh(K_2 d_2)}, \tag{S23}$$

$$K_2^2 = k^2 + i\mu_0 \sigma_2 \omega, \tag{S24}$$

$\sigma_2$ is the conductivity of the nonmagnetic metallic layer located above the ferromagnetic one ($L<y<L+d_2$),

$$S_j = \frac{ik \exp(-d_s |k|)}{K_2 \sinh(K_2 d_2) + |k| \cosh(K_2 d_2)} j_{sk}, \tag{S25}$$

$j_{sk}$ is the Fourier-image of the microwave current density in the spin-wave antenna that drives the magnetization precession, and $d_s$ is the thickness of the dielectric spacer sandwiched between the surface of the upper metallic nonmagnetic layer and the antenna.

$$C_{hxi} = -\frac{-\alpha\omega_M K^2 q_i^4 + \omega q_i^3 + A_1 q_i^2 + \omega K^2 q_i + K^2 A_2}{(K^2 - q_i^2) A_{3i}}, \tag{S26}$$

$$A_1 = \omega_H K^2 - \alpha\omega_M \left(|K|^4 - 3k^2 K^2\right), \tag{S27}$$

$$A_2 = -K^2 \omega_H - \omega_M \alpha k^2 \bar{K}^2 + \omega_M \omega \mu_0 \sigma, \tag{S28}$$

$$A_{3i} = \omega_H \left(q_i^2 - K^2\right) + \omega_M \left(q_i^2 - i\omega\mu_0\sigma\right) - \alpha\omega_M (q_i^2 - k^2)(q_i^2 - K^2), \tag{S29}$$

and,

$$C_{myi} = i\frac{\omega_M q_i k + \omega(K^2 - q_i^2)}{-\omega_H(K^2 - q_i^2) + \omega_M \left(q_i^2 - i\omega\mu_0\sigma\right) + \omega_M \alpha \left(q_i^2 - k^2\right)(K^2 - q_i^2)}. \tag{S30}$$

The Fourier-image of the microwave electric field $e_{zk}$ ($y_0$) (Eq. (22)) at the height of the antenna ($y_0=L+d_2+d_s$) has two contributions:

$$e_{zk} = e_{zk\,ind} + e_{zk\,m}. \tag{S31}$$

The $e_{zk\,ind}$ contribution is independent of the dynamic magnetization. This contribution is the electric field induced by the current in the antenna in the absence of magnetization precession and is always present. The expression for $e_{zk\,ind}$ reads,

$$e_{zk\,ind}(y_0) = i\omega\mu_0 \frac{k^2 \cosh(K_2 d_2)\cosh(|k|d_s) + K_2 \sinh(K_2 d_2)\sinh(|k|d_s)}{|k|\left[k^2 \cosh(K_2 d_2) + K_2 |k| \sinh(K_2 d_2)\right]} \exp(-|k|d_s) j_{sk}. \tag{S32}$$

As this contribution is present off the spin-wave band, it can be used to calculate the characteristic linear inductance of the coplanar line when it is not loaded by the excitation of spin waves. The method to calculate it is analogous to the method of calculating the linear capacitance (see Section 9) and is based on a condition that the total current through a coplanar line must vanish. This condition is dual to the condition (Eq. (S53)) from Section 9 for the total linear charge of the coplanar line.

For $|k| \gg 1/d_s$, $1/d_2$, $(\mu_0 \sigma_2 \omega)^{1/2}$, Eq. (S32) reduces to

$$e_{zk\ ind\infty}(y_0) = \frac{i\omega\mu_0}{2|k|} j_{sk} . \tag{S33}$$

The second contribution is the electric field $e_{zk\ m}$ induced by the precessing magnetization vector. This contribution is significant on the spin-wave frequency band only. It can be expressed through the $x$-component of the microwave magnetic field $h_{xk}$ of the precessing magnetization at the surface of the ferromagnetic layer that faces the antenna ($y=L$):

$$e_{zk\ m}(y_0) = i\omega\mu_0 |k| \frac{h_{xk}(y=L)}{k^2 \cosh(K_2 d_2) + K_2 |k| \sinh(K_2 d_2)} \exp(-|k|d_s) . \tag{S34}$$

The magnetic field at the surface is obtained from Eq. (21):

$$h_{xk}(y=L) = \sum_{i=1}^{6} C_{hxi} M_i^y \exp(q_i L) . \tag{S35}$$

Note that Eq. (S35) is a linear function of $j_{sk}$, because of a linear dependence of $M_i^y$ on it. For $j_{sk}=1$, the Fourier transform of the sum of Eq. (S32) and Eq. (S34) represents the Green's function of the electric field $E_z^0(s)$ Eq. (22). This Fourier transform has to be evaluated numerically and requires significant computer time to complete. To accelerate the numerical calculation, one can simplify the contribution of Eq. (S32) to the Green's function. This is possible because the integral,

$$\frac{1}{\pi} \int_{k_c}^{\infty} \frac{i\omega\mu_0}{2|k|} \cos(ks) ds = -\frac{i\omega\mu_0}{2\pi} C_i(k_c s) . \tag{S36}$$

where $k_c$ is some arbitrary wavenumber satisfying the condition $k_c \gg 1/d_s$ and $C_i(x)$ is the cosine integral function, for which an infinite-fraction representation exists. Thus, Eq. (22) reduces to

$$E_z^0(s) = \frac{1}{2\pi} \int_0^{\infty} \left[ e_{zk\ m}(k) \exp(-iks) + e_{zk\ m}(-k) \exp(iks) \right] dk + \frac{1}{\pi} \int_0^{k_c} e_{zk\ ind}(k) \cos(ks) dk - \frac{i\omega\mu_0}{2\pi} C_i(k_c s) , \tag{S37}$$

where $e_{zk\ m}$ and $e_{zk\ ind}$ are defined by Eq. (S32) and Eq. (S34) and under assumption $j_{sk}=1$.

## 13. Derivation of the expression for the capacitance of the spin-wave antenna

We will carry out the derivation in the electrostatic approximation. Therefore, no currents will be included in the model, and the upper surface ($y=L+d_2$) of the stack of the three metal layers can be considered as a surface of an ideal metal. Thus, de-facto, we are dealing with a capacitance of a backed coplanar line. As before, we assume that the coplanar antenna has zero thickness in the direction $y$. The antenna is characterized by a charge density distribution $\rho(x)$ along the antenna cross-section. In the $z$-direction, the charge density is distributed homogeneously. Therefore, $\rho(x)$ represents an area charge density (measured in C/m$^2$). Accordingly, the voltages and capacitances, with which we will be dealing below, are linear voltages and capacitances (i.e. per unit length of the antenna in the direction $z$).

For an electrostatic field in an isotropic dielectric medium without charges we have

$$\nabla \times \mathbf{E} = 0, \tag{S38}$$

$$div(\mathbf{E}) = 0. \tag{S39}$$

In the Fourier space, these expressions reduce to an equation as follows

$$d^2/dy^2(E_{yk} - k^2) = 0. \tag{S40}$$

Its solution for the area inside the insulating spacer ($L+d_2 < y < L+d_2+d_s$) reads

$$E_{yks} = A_s \exp(ky) + B_s \exp(-ky). \tag{S41}$$

For the area (Area 3) above the antenna ($y > L+d_2+d_s$), the solution is trivial

$$E_{yk3} = B_3 \exp(-ky). \tag{S42}$$

From Eq. (S39) we can also obtain a relation between $E_{yk}$ and $E_{xk}$. It reads,

$$E_{xk} = -\frac{i}{k} dE_{yk}/dy. \tag{S43}$$

The electrostatic boundary condition for the surface of the ideal metal ($y=L+d_2$) is trivial

$$E_{xks}(y = L+d_2) = 0. \tag{S44}$$

The respective boundary conditions at the antenna are more complicated, as they include the area charge density:

$$E_{xks}(y = L+d_2+d_s) = E_{xk3}, \tag{S45}$$

$$\varepsilon_0 E_{yk3}(y=L+d_2+d_s) - \varepsilon_0\varepsilon_s E_{yks}(y=L+d_2+d_s) = \rho_k. \tag{S46}$$

Where $E_{xk3}$ is the x-component of the electric field above the antenna (Area 3), $\varepsilon_s$ is the relative dielectric permittivity of the spacer, and $\rho_k$ is the Fourier-image of $\rho(x)$. (We model the antenna as a surface charge density.)

Application of the boundary conditions yields expressions for the coefficients of Eq. (S41) and Eq. (S42). They are as follows:

$$A_s = T\exp[-|k|(L+d_2)], \quad B_s = T\exp[|k|(L+d_2)], \tag{S47}$$

where,

$$T = \frac{\rho_k}{\varepsilon_0} \frac{2}{\sinh(|k|d_S) + (1+\varepsilon_s)\cosh(|k|d_S)}. \tag{S48}$$

Integrating $E_{yks}$ over the thickness of the insulating spacer yields the Fourier image of the voltage $V_k$ across the spacer thickness,

$$V_k = \rho_k K_V, \tag{S49}$$

where,

$$K_V = -\frac{1}{\varepsilon_0 |k|} \frac{\sinh(|k|d_S)}{\sinh(|k|d_S) + \varepsilon_s \cosh(|k|d_S)}, \tag{S50}$$

Equation (S49) demonstrates that,

$$G_V(s) = \frac{1}{2\pi} \int_{-\infty}^{\infty} K_V \exp(-iks)dk, \tag{S51}$$

represents a Green's function of the electric voltage. Then, a voltage induced by an arbitrary charge density $\rho(x)$ is given by,

$$V(x) = \int_{-\infty}^{\infty} G_V(x-x')\rho(x')dx'. \tag{S52}$$

One sees that $K_V$ (Eq. (S50)) is an even function of $k$. Therefore, Eq. (S51) reduces to,

$$G_V(s) = \frac{1}{\pi} \int_{-\infty}^{\infty} K_V \cos(ks)dk. \tag{S53}$$

For $|k|d_s \gg 1$, Eq. (S50) reduces to,

$$K_V(|k|d_s \gg 1) = -\frac{1}{\varepsilon_0 |k|} \frac{1}{1+\varepsilon_s} . \tag{S54}$$

This allows one to obtain a closed-form expression for the contribution of the large Fourier wavenumbers to the integral in Eq. (S53)

$$G_{V\infty}(s) = \frac{1}{\pi} \int_{k_c}^{\infty} K_V \cos(iks)\, ds = \frac{1}{\pi\varepsilon_0(1+\varepsilon_s)} C_i(k_c s) , \tag{S55}$$

where $k_c$ is some arbitrary wavenumber satisfying the condition $k_c \gg 1/d_s$, and $C_i(x)$ is the cosine integral function. Thus, Eq. (S53) reduces to

$$G_V(s) = \frac{1}{\pi} \int_0^{k_c} K_V \cos(ks)\, dk + \frac{1}{\pi\varepsilon_0(1+\varepsilon_s)} C_i(k_c s) . \tag{S56}$$

This expression can easily be evaluated numerically.

The electric voltage must be uniform at the surface of an ideal metal. This yields an integral equation for $\rho(x)$,

$$V_i = \sum_{n=1}^{3} \int_{(x_n)} G_V(x-x')\rho(x')\, dx' \quad . \tag{S57}$$

where $V_i$ ($i=1,2,3$) are voltages for individual strips of the coplanar line.

Our ultimate goal is to find the linear capacitance of the coplanar line. The capacitance can be defined as,

$$C = Q_1/V_{12}, \tag{S58}$$

where,

$$V_{12} = V_1 - V_2, \tag{S59}$$

is the potential difference between the signal line (the central strip) and one of the ground lines (a side strip), and

$$Q_1 = \int_{(x_1)} \rho(x')\, dx' \quad . \tag{S60}$$

is the total linear charge for the same line. This expression holds, provided the total linear charge for the whole coplanar line is zero:

$$Q_1 + Q_2 + Q_3 = 0. \tag{S61}$$

For symmetry reasons we then have,

$$Q_2 + Q_3 = -Q_1/2. \tag{S62}$$

To be able to make use of this condition, we solve the integral equation (Eq. (S57)) with respect to $\rho(x)$ numerically three times. First, we set $V_1=1$ V/m and $V_2=V_3=0$. The solution yields,

$$Q^{(1)} = Q_1 + Q_2 + Q_3 = \sum_{n=1}^{3} \int_{(x_n)} \rho(x')\,dx' \quad . \tag{S63}$$

Then we repeat the same procedure for $V_1=0$, $V_2=V_3=1$ V/m to obtain,

$$Q^{(2)} = \sum_{n=1}^{3} \int_{(x_n)} \rho(x')\,dx' \quad , \tag{S64}$$

in the same way. For the last numerical solution, we use $V_1=1$ V/m and $V_2=V_3=-V_1 Q^{(1)}/Q^{(2)}$. These new values of $Q_1$, $Q_2$, and $Q_3$ satisfy Eq. (S61), and we can now use Eq. (S58) to calculate the capacitance. To this end, it is convenient to reduce Eq. (S58) to,

$$C = \frac{Q^{(2)}}{Q^{(2)} + Q^{(1)}} \frac{Q_2 + Q_3}{V_1} . \tag{S65}$$

Numerical calculations employing Eq. (S65) demonstrate good agreement with the known analytical formula [34] that exists for the case of an unbacked coplanar line ($d_s=\infty$). Furthermore, if one calculates the linear inductance for the coplanar line using $e_{zk\,ind}$ (see Section 8), from the two values, one obtains a value of the characteristic impedance that is very close to [34].

For the general case of a backed coplanar line, there are no analytical expressions for $C$ and the linear inductance, and one has to use the formalism above to calculate the capacitance.

Note that C given by Eq. (S65) is the full capacitance of the CPL. From Figure 3 in the main text, it is clear that it represents the parallel connection of two capacitances giving rise to two admittances $Y_{12}$ and $Y_{13}$. This implies that $Y_{12}=Y_{13}=i\omega C/2$.

## 14. Details of the derivation of the integral equation Eq. (25), and the expression for the matrix of inductive impedances Eq. (29)

We employ a model based on Kirchhoff's voltage law in order to formulate an integral equation that is the central point of the self-consistent solution. Consider an equivalent circuit for the unit length of a strip of the antenna shown in Figure S6. The strip is of an infinite length in the direction $z$, and we assume translational invariance in this direction (i.e. all relevant quantities do not depend on $z$). We break the strip width $w_s$ into elementary sections of width $dx$. Each section is characterized by a current $j(x)dx$. An external voltage $E_{ext}$ is applied to the section of unit length. Note that because this is a voltage per unit length – "linear voltage", it has units of V/m, i.e. the same as for an electric field, therefore we use the letter $E$ to denote it. It is important to understand that the linear voltage is applied *uniformly* across the strip width, as expected for an external voltage and the translational invariance.

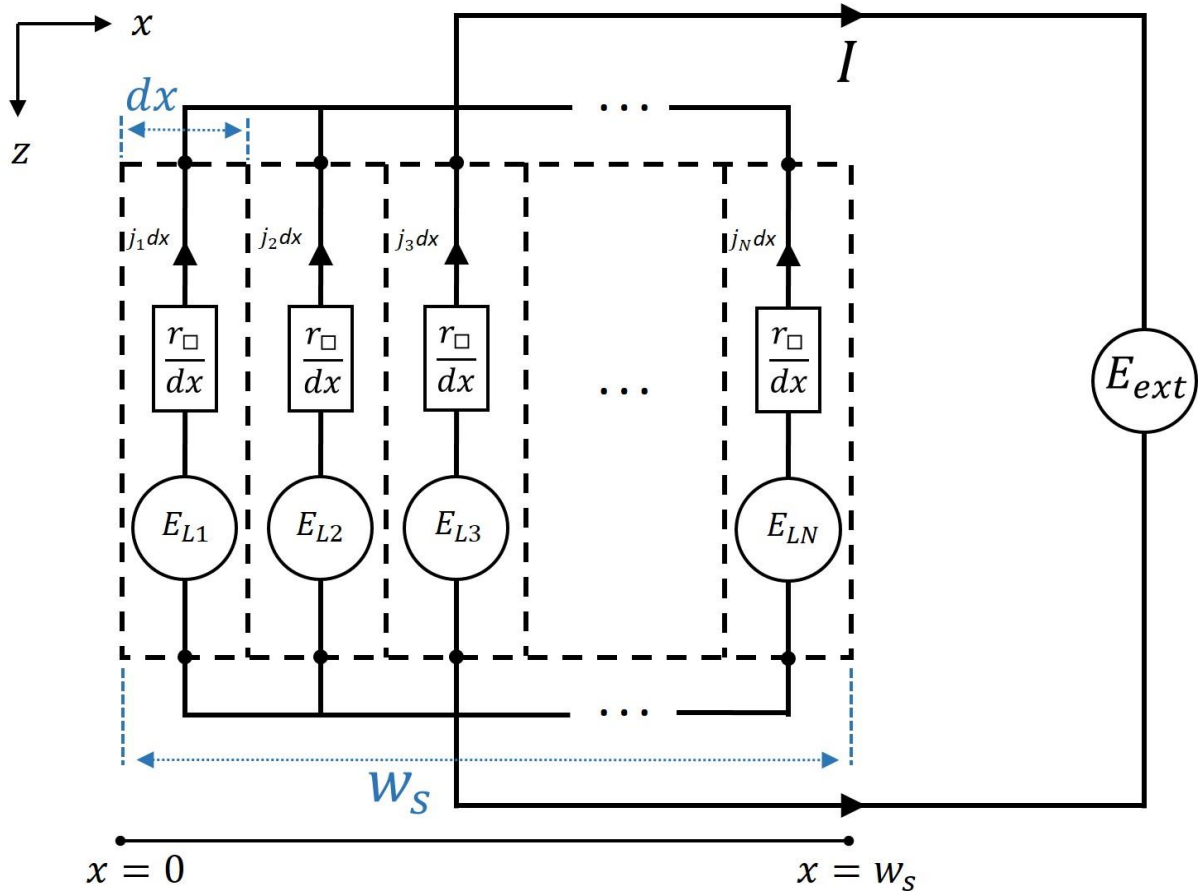

*Figure S6: Equivalent circuit for a single-strip input antenna of width $w_s$.*

Each elementary section $i$, where $i=1, 2, \ldots N$ and $N = w_s/dx$, is characterized by a linear ohmic voltage drop $\rho/(t_a dx) \, j_i dx = r_\square \, j_i$ and a linear voltage $E_{Li}$ induced in it due to self-inductance of the elementary section and mutual inductance to which all other elementary sections contribute. We will call $E_L$ a Faraday-induction voltage in the following. Here $r_\square = \rho/t_a$ is the sheet antenna resistance, $\rho$ is the resistivity of the antenna metal, and $t_a$ is the CPL thickness in the direction $y$. Hence, we have a parallel connection of $N$ circuits, each of which consists of a resistor $r_\square/dx$ and a local voltage source $E_{Li}$ which is different for each $i$. Each of these circuits is connected to a global voltage source $E_{ext}$, thus forming a closed loop (See Figure S6). Following the Kirchhoff law, for the $i$-th elementary section we have $-E_{ext} + r_\square j_i + E_{Li} = 0$. This yields an equation $r_\square j(x) + E_L(x) = E_{ext}$, where we replaced $j_i$ with $j(x)$ for the respective $x$ and $E_{Li}$ with $E_L(x)$ for the same $x$. ($E_L(x)$ is given by the continuous Green's function of the electric field - Eq. (23)). The Green's function accounts for both intrinsic inductance of the CPL line (that does not vanish off the spin-wave band) and the excitation of spin waves on the band. Furthermore, we need to take into account that not only the strip itself, but the other strips of the CPL contribute to $E_L$. This yields an integral equation as follows

$$E_m = r_\square j_m(x) + \sum_{n=1}^{3} \int_{(x_n)} G_E(x-x') j_n(x') dx' \quad , \tag{S66}$$

where $E_m$ is the external linear voltage applied to the $m$-th strip of the antenna (the ground or a signal line), $j_m$ is the current density in the $m$-th strip, and $(x_n)$ means integration over the width of the $n$-th strip of the CPL, and $m,n=1,2,3$.

As an external voltage, $E_i$ is assumed to be uniform across the width of each individual strip (see the discussion above), but $E_i$-values are different for different strips in the general case.

From this equation, one sees that the uniform external voltage results in a non-uniform current density $j_n(x)$. This is a consequence of the current crowding effect. In the absence of inductance, the current density is uniform, $j(x) = E_{ext}/r_\square$ because Ohm's Law is local. Then, following the Kirchhoff law, the voltage drop due to the *local* ohmic losses perfectly compensates the external voltage ($j(x) r_\square - E_{ext}=0$). The Faraday induction that underlies the inductance is not local, therefore, if the conductor's inductance is non-vanishing, a current flowing elsewhere can contribute to the compensation of an external voltage at any point of the cross-section of the conductor. This is the essence of the Skin Effect - a current flowing near the surface of a bulk piece of metal can perfectly compensate an external voltage applied uniformly to the specimen. Note that the external voltage is perfectly compensated *everywhere* inside the sample, not only where the current flows. In the case of our two-dimensional geometry of a strip of infinitely small thickness ($t_s=0$), the skin effect results in current density localization near the strip edges provided the strip width is large enough. If the strip width is small, we expect ohmic losses to dominate over inductance (the first term on the r.h.s. of Eq. (S66) to be much larger than the second one), and, consequently, the microwave current density to be uniform over the strip width [35].

Note that our formulation of the integral equation is in agreement with the established one [24]. In those early publications, ohmic losses were ignored while writing down the integral equation. This implies that the metal from which a spin-wave antenna is made was considered as ideal ($r_\square=0$). For ideal metal, the component $B_n$ of the magnetic induction perpendicular to the metal surface must vanish. This yields the integral equations for current density from those papers. Now recall that the drop of the external linear voltage has units of the electric field, and represents the external electric field applied to the stripline. Then, for the field **E** inside the metal, we may write

$$\mathbf{E} = i\omega \mathbf{A} - \mathrm{grad}(\varphi) \ . \tag{S67}$$

In our case, $\text{grad}(\varphi) = E_{ext}$ is the gradient of the external voltage applied to the strip. The gradient equals the linear voltage $E_{ext}$. $\mathbf{A}$ is the vector potential of the magnetic field. It accounts for the electric field due to the strip inductance. The condition $B_n = B_y = 0$ implies that

$$\partial \mathbf{A} / \partial x - \partial \mathbf{A} / \partial z = 0. \tag{S68}$$

In addition, translational invariance in the $z$-direction requires that $\partial \mathbf{A} / \partial z = 0$. This reduces Eq. (S68) to $\mathbf{A}(x) = const(x)$. Given that $i\omega \mathbf{A}(x)$ is the electric field generated due to the strip inductance (Faraday-induction field) we see that the requirement $B_y = 0$ is equivalent to a requirement of uniformity of the Faraday-induction field across the strip width. Being uniform, the Faraday-induction field can perfectly compensate the external electric field such that $i\omega \mathbf{A}(x) - E_{ext} = 0$ at any point of the strip, as required by both the condition for the electric field inside the ideal metal ($\mathbf{E}=0$ in Eq. (S67)) and the Kirchhoff law. The current that is responsible for creating the Faraday-induction field is the strip edge current, and as such, it is highly non-uniform along $x$.

This demonstrates that the integral equation (Eq. (S66)) reduces to the integral equations for the current density from [24] in the case $r_\square=0$.

Let us now figure out how to convert the solution of Eq. (S66) into antenna impedance. Consider first a single-strip spin-wave antenna of a width $w_s$. The power radiated by an antenna is given by [24]

$$\frac{1}{2} Z |I|^2 = \frac{1}{2} \int_{(w_s)} \overline{j}(x) e_z(y_0, x)\, dx, \tag{S69}$$

where $Z$ is the linear complex-valued radiation impedance of the antenna, $I$ is the total current in the antenna, $j(x)$ is the density of the current

$$I = \int_{(w_s)} j(x)\, dx, \tag{S70}$$

and the dash above $j(x)$ denotes complex conjugation. The left-hand side of the equation represents microwave power absorbed by a unit of antenna length. Hence, on the right-hand side of the equation, we must have the same absorbed power, but $x$-co-ordinate resolved. In the original paper [24], an antenna made of ideal metal was considered, therefore $e_z(y_0, x)$ was the electric field induced by the excited spin waves inside the antenna. This is the only contribution to $E_{ext}$ in the case of an ideal metal, as already mentioned above. In our case of $r_\square \neq 0$, the absorbed power consists of ohmic losses and inductance, the latter being a sum of microwave power converted into the power of spin waves and off-band antenna inductance. Hence, we must have both ohmic-loss and Faraday-inductance contributions on the right-hand side. However, following Eq. (S66), their sum must be equal to $E_{ext}$. This implies that in our case, Eq. (S69) converts to

$$Z|I|^2 = E_{ext} \int_{(w)} \bar{j}(x)\, dx, \quad (S71)$$

where we pulled $E_{ext}$ out of the integral because we know that it must be uniform over the strip width. Given Eq. (S70), the integral on the right-hand side of this equation is just $\bar{I}$, and Eq. (S71) reduces to

$$Z = E_{ext}/I. \quad (S72)$$

This equation allows one to calculate the antenna linear impedance, once the integral equation (Eq. (S66)) has been solved.

Recall that Eq. (S72) is valid for a single-strip antenna, i.e. a microstrip antenna. For a coplanar-line antenna, each strip is characterized by its own value of $E_{ext}$. This implies that $E_{ext}$ must now represent a vector $\mathbf{E}_v=(E_1, E_2, E_3)$ where 1, 2, and 3 are the strip numbers. Similarly, the scalar linear impedance $Z$ converts into a matrix $\mathbf{Z}$ with elements $Z_{ij}$. The diagonal elements $Z_{ij}$ of the matrix represent self-impedances of the strips, and the non-diagonal ones their mutual impedances. Then, the matrix-vector version of Eq. (S72) reads

$$\mathbf{Z} = \mathbf{I}^{-1}\mathbf{E}_v. \quad (S73)$$

where $\mathbf{I}^{-1}$ is a 3x3 matrix with elements given by Eq. (27) in Section 2.2.

## 15. Details of the derivation of the expression for the electromotive forces $\Delta U_i$ induced in the output antenna

The electric field of the incident spin wave $E_{inc}(x)$ generates an electro-motif forces (e.m.f.) $\Delta U_i$ in the metal strips of the output antenna. Recall that the wave's electric field has only one component – the $z$-component that is along the antenna strips. It is important to understand that there is a difference between the linear e.m.f., (i.e. e.m.f per unit length of the antenna) $\Delta U_i$ and $E_{inc}(x)$. The field $E_{inc}(x)$ has a form of a traveling wave, but the e.m.f. must obey the same requirement of uniformity of voltage across the widths of the metal strips of the antenna. This suggests that there must be a current flowing locally at each $z$ cross-section of each antenna strip that redistributes electric charges in the direction of the strip width (the $x$-direction). The local current makes the total electric field within the antenna width (the field is the same as the linear e.m.f. $\Delta U_i$) uniform within the strip width. We will term this current a local current; its density is $j_l(x)$.

For simplicity, let us consider a single-strip antenna first. The idea above can be cast in an integral equation as follows

$$\Delta U = E_{inc}(x) + r_\square j_l(x) + \int\limits_{(w)} G_E(x-x')j_l(x')dx', \tag{S74}$$

where $G_E(x)$ is the same Green's function of the electric field of a spin wave as before Eq. (22), $w$ is the width of the single strip and $\Delta U$ is the e.m.f. The e.m.f. is uniform within the width of the strip, hence we have a constant left-hand side of the equation. Introduce an operator with a kernel $G^{-1}(x-x')$ as an operator that is inverse to the integral operator with the kernel $G(x-x')$ of the integral on the right-hand side of the equation (S74)

$$G(x-x') = r_\square \delta(x-x') + G_E(x-x')'. \tag{S75}$$

Here $\delta(x)$ is the Kronecker Delta function, which we used to add the Ohmic-loss term to the kernel. This implies that $G^{-1}(x-x')$ satisfies the condition as follows

$$\int\limits_{(w)} G^{-1}(x-x')' \left[ \int\limits_{(w)} G(x'-x'')dx'' \right] dx' = 1. \tag{S76}$$

Introduce a short-hand notation for the operators such that Eq. (S76) reads

$$G^{-1} \otimes G \otimes 1 = 1. \tag{S77}$$

where 1 on the left-hand side is just a constant 1. It was added for consistency of the expression.

Assume that the operator $G^{-1} \otimes$ is known and apply it to both sides of Eq. (S74). We obtain

$$\Delta U\, G^{-1}(x-x') \otimes 1 = G^{-1}(x-x') \otimes E_{inc}(x') + j_l(x). \tag{S78}$$

Recall that $\Delta U$ is a constant. Therefore, it can be pulled out of the operator, as we did above.

By integrating the density of the local current over the strip width $w$ we obtain the total local current. We do not expect the total local current to contribute to the density $j_g$ of the global current $I_g$. The global current is a current that is continuous along the whole length of the antenna and obeys the Telegrapher Equations (Eqs. (30)). There is only one way to satisfy these requirements. This is by requiring that the total local current vanishes:

$$\int\limits_{(w)} j_l(x)\, dx = 0. \tag{S79}$$

Given that, integrate Eq. (S78) over the strip width $w$. We obtain

$$\Delta U \int\limits_{(w)} dx\, G^{-1}(x-x') \otimes 1 = \int\limits_{(w)} dx\, G^{-1}(x-x') \otimes E_{inc}(x'), \tag{S80}$$

where we used the condition of Eq. (S79) to get rid of the second term on the right-hand side of Eq. (S78). Now, we solve Eq. (S80) for $\Delta U$ to obtain

$$\Delta U = \frac{\int\limits_{(w)} dx \int\limits_{(w)} G^{-1}(x-x')E_{inc}(x')dx'}{\int\limits_{(w)} dx \int\limits_{(w)} G^{-1}(x-x')dx'} \ . \tag{S81}$$

Let us now analyze Eq. (S81). As follows from Eqs. (S74) and (S76), the quantities on the left- and right-hand sides of Eq. (S80) have units of current. This suggests that the quantities represent currents that are induced by sources of linear voltage $E_{inc}(x)$ and $1(x)$ respectively (where $1(x)$ denotes a uniform current density with a unit amplitude). Accordingly, $G^{-1}(x-x')\otimes 1$ represents the density of a current $j_1(x)$ induced by the source $1(x)$. Then from the definition of the inverse operator $G^{-1}\otimes$(Eqs. (S75)-(S76)-), we find that $j_1(x)$ satisfies the integral equation

$$\int\limits_{(w)} \left[ r_\square \delta(x-x') + G_E(x-x') \right] j_1(x')dx' = 1. \tag{S82}$$

which is the one-strip analog of Eq. (25) written for the linear-voltage amplitude of 1. The denominator of Eq. (S81) contains an integral of $j_1(x)$, which is a total current associated with $j_1(x)$. Given the analogy of Eq. (S82) to Eq. (25), it is now easy to realize that the denominator of Eq. (S81) represents $Z^{-1}$, where Z is the linear inductive impedance of the strip. Z is the single-strip analog of the matrix **Z** (Eq. (29)).

Similarly, we can introduce a current density

$$j_{inc}(x) = \int\limits_{(w)} G^{-1}(x-x')E_{inc}(x')dx' \ . \tag{S83}$$

for the numerator of Eq. (S82). Then the numerator represents a total current $I_{inc}$ associated with $j_{inc}(x)$

$$I_{inc} = \int\limits_{(w)} j_{inc}(x)dx \ . \tag{S84}$$

Furthermore, similar to Eq. (S82), instead of using Eq. (S83) we may employ an integral equation as follows to define $j_{inc}(x)$

$$\int\limits_{(w)} \left[ r_\square \delta(x-x') + G_E(x-x') \right] j_{inc}(x')dx' = E_{inc}(x) \ . \tag{S85}$$

We will explain the importance of Eq. (S85) later on.

Finally, we can now rewrite Eq. (S81) in a very compact and physically transparent form

$$\Delta U = ZI_{inc}. \qquad (S86)$$

This equation shows that the e.m.f. induced by a spin wave incident on a metallic strip is given by the product of the linear inductance of the antenna and some current $I_{inc}$. It is important to understand that this current is not the current that flows in the real antenna. The current in the antenna would be equal to $I_{inc}$ if the antenna was much shorter than the electromagnetic wave induced in the strip for the given frequency (which is our case) and if it was terminated by a perfectly matched load at its far end. In our case of the short-ended antenna, we expect the formation of a wave reflected from the short end. Then the actual current is obtained by solving Telegrapher Equations that include $\Delta U$ and the linear capacitance of the strip [28].

Note that our theory of spin-wave excitation shows that the inductance Z includes several contributions – the intrinsic off-spin-wave-band inductance of the strip, a contribution from the Ohmic resistance of the strip $r_\square$, and a contribution of spin-waves to the inductance. The latter waves are secondary waves excited by the strip when a current is induced in it by the incident spin wave. It is clear that two secondary waves are excited – in the positive and negative directions of the axis *x*. The secondary wave that is co-propagating with the incident one, must interfere destructively with the latter, in order to reduce the spin wave amplitude past the antenna. This ensures the balance of power – power must be taken from the incident spin wave to form a microwave signal in the antenna. The second secondary wave represents a wave reflected from the output antenna - it propagates backward (that is toward the input antenna) thus forming a partial standing wave in front of the output antenna. In this way, the process of reception of a microwave signal by the antenna physically represents a process of scattering of a spin wave from a metallic strip [28]. Because the process of excitation and reception of MSSW is highly non-reciprocal, one may expect that the backward scattered field for a spin wave incident onto the strip from the direction of more efficient wave excitation is small. This is in agreement with the asymmetry of the matrix **Z** (Eq. (29)).

Now let us return to the importance of Eq. (S85). A numerical model that will solve Eq. (S85) will convert the integral kernel into a matrix $\Gamma$ and the functions $j_{inc}(x)$ and $E_{inc}(x)$ into vectors that contain values of these functions at discrete points $x_i$ of a mesh. The integral equation then converts into an inhomogeneous system of linear equations. Numerical methods of linear algebra allow fast and very accurate solution $j_{inc}(x_i)$ of the system. Conversely, finding the inverse operator $G^{-1}\otimes$ (Eq. (S77)) numerically requires inverting the matrix $\Gamma$. Inverting a matrix numerically is a much slower and much less accurate process than solving numerically a system of linear equations. In addition, the matrix $\Gamma$ is the same matrix used to calculate Z. This makes the numerical implementation of the solution of Eq. (21) even faster by re-using the matrix that has already been built while computing Z.

The last step of our derivation is extending Eq. (S83) to the case of the coplanar line. This is straightforward. Equations (42)-(44) in the main text show the solution.